\documentclass[11pt]{article}
\usepackage{a4wide,amssymb,latexsym,epsfig,amsmath,amsfonts,oldgerm,array,graphicx}
\usepackage[english]{babel}
\renewcommand{\thesection}{\arabic{section}}
\newcommand{\app}[1]{ 
     \renewcommand{\thesection}{#1} 
     \section*{Appendix #1}\setcounter{equation}{0}} 

\numberwithin{equation}{section}

\newcommand{\be}{\begin{equation}}
\newcommand{\ee}{\end{equation}}
\newcommand{\ba}{\begin{array}}
\newcommand{\ea}{\end{array}}
\newcommand{\bea}{\begin{eqnarray}}
\newcommand{\eea}{\end{eqnarray}}
\newtheorem{Theorem}{Theorem}

\newcommand{\al}{\alpha}

\newcommand{\ga}{\gamma}
\newcommand{\Ga}{\Gamma}
\newcommand{\de}{\delta}
\newcommand{\D}{\Delta}
\newcommand{\ep}{\epsilon}

\renewcommand{\th}{\theta}

\newcommand{\ka}{\kappa}
\newcommand{\la}{\lambda}

\newcommand{\laii}{\lambda_{\rm\scriptscriptstyle I\kern-.1em I}}
\newcommand{\rh}{\varrho}

\newcommand{\pa}{\partial}
\newcommand{\R}{{\rm I\kern-0.2em R}}
\newcommand{\C}{{\rm C\kern-0.6em I}}
\newcommand{\N}{{\rm I\kern-0.2em N}}
\newcommand{\Z}{{\sf Z\kern-0.4em Z}}

\newcommand{\ti}{\times}

\newcommand{\na}{\nabla}

\renewcommand{\le}{\left}
\newcommand{\ri}{\right}
\newcommand{\ra}{\rightarrow}

\newcommand{\xx}{{\bf x}}

\newcommand{\vv}{{\bf v}}

\newcommand{\rrho}{\mbox{\boldmath$\rho$}}
\newcommand{\BB}{{\bf B}}
\renewcommand{\AA}{{{\bf B}_a}}
\newcommand{\MM}{{{\bf B}_m}}
\newcommand{\EE}{{\bf E}}
\newcommand{\JJ}{{\bf J}}
\newcommand{\eee}{{\bf e}}
\newcommand{\bb}{{\bf b}}
\newcommand{\nn}{{\bf n}}

\newcommand{\rd}{{\rm d}}
\newcommand{\SSS}{S\! S}
\newcommand{\SC}{S\! C}
\newcommand{\SCh}{\widehat{S\! C}}
\newcommand{\Vk}{V\!\!\kappa}
\newcommand{\Vkti}{V\!\tilde{\kappa}}
\newcommand{\Ver}{\widetilde{V}\!\!\eta\!\varrho}

\newcommand{\cM}{{\cal M}}

\newcommand{\cA}{{\cal A}}

\newcommand{\dti}{\widetilde{d}}
\newcommand{\kti}{\widetilde{\kappa}}
\newcommand{\Cti}{\widetilde{C}}

\newcommand{\Rti}{\widetilde{R}}

\newcommand{\Bh}{\widehat{B}}
\newcommand{\Gh}{\widehat{G}}
\newcommand{\Ch}{\widehat{C}}
\newcommand{\Mh}{\widehat{M}}
\newcommand{\Go}{\overline{G}}
\newcommand{\Mu}{\underline{M}}
\newcommand{\Mo}{\overline{M}}
\newcommand{\Pu}{\underline{P}}
\newcommand{\Po}{\overline{P}}
\newcommand{\Bo}{\overline{B}}
\newcommand{\Co}{\overline{C}}
\newcommand{\ro}{\overline{\varrho}}
\newcommand{\ru}{\underline{\varrho}}
\newcommand{\lao}{\overline{\lambda}}
\newcommand{\lau}{\underline{\lambda}}
\newcommand{\muo}{\overline{\mu}}
\newcommand{\muu}{\underline{\mu}}

\newcommand{\kiiq}{\overline{\kappa}_{\rm\scriptscriptstyle I\kern-.1em I}}
\newcommand{\kiiiq}{\overline{\kappa}_{\rm\scriptscriptstyle
I\kern-.1em I\kern-.1em I}}

\newcommand{\kii}{\kappa_{\rm\scriptscriptstyle I\kern-.1em I}}

\newcommand{\II}{{I\kern-.1em I}}
\newcommand{\III}{{I\kern-.1em I\kern-.1em I}}

\newcommand{\ii}{{\rm\scriptscriptstyle I\kern-.1em I}}
\newcommand{\iii}{{\rm\scriptscriptstyle I\kern-.1em I\kern-.1em I}}

\newcommand{\noi}{\noindent}

\newcommand{\ds}{\displaystyle}
\begin{document}
\title{The axisymmetric antidynamo theorem revisited}
\author{Ralf Kaiser${}^+$ and Andreas Tilgner${}^\#$\\[2ex]
         \small ${}^+$ Fakult\"at f\"ur Mathematik und Physik,
          Universit\"at Bayreuth, D-95440 Bayreuth, Germany\\[1ex]
     \small ${}^\#$ Institut f\"ur Geophysik, Universit\"at G\"ottingen, D-37077 G\"ottingen, Germany\\[1ex]
     \small ralf.kaiser@uni-bayreuth.de, andreas.tilgner@geo.physik.uni-goettingen.de} 
\date{July, 2013}
\maketitle

%
%
\begin{abstract}
The axisymmetric kinematic dynamo problem is reconsidered and a number of open questions are answered. Apart from axisymmetry and smoothness of data and solution we deal with this problem under quite general conditions, i.e.\ we assume a compressible fluid of variable (in space and time) conductivity moving in an arbitrary (axisymmetric) domain. We prove unconditional, pointwise and exponential decay of magnetic field and electric current to zero. The decay rate of the external (meridional) magnetic field can become very small (compared to free decay) for special flow fields and large magnetic Reynolds numbers. We give an example of that. On the other hand, we show for fluids 
with weak variation of mass density and conductivity that the meridional and azimuthal decay rates do not drop significantly below those of free decay. 

\vspace{5mm}

\noi Key Words: Magnetohydrodynamics, dynamo theory, axisymmetric theorem.
\end{abstract}
\section{Introduction}
The axisymmetric (or Cowling's) theorem is the oldest and most renowned of all antidynamo theorems which generally preclude under certain assumptions the maintenance of the magnetic field against ohmic loss by the dynamo process. Here, the assumption is axisymmetry (of magnetic field, flow field, conductivity distribution and shape of the conductor). This theorem had considerable significance in the early days of dynamo theory, where convincing analytical arguments, computational or even experimental evidence for the dynamo process were yet far away. In view of nearly axisymmetric magnetic fields of the earth and the sun this theorem called in question the dynamo process as such. No wonder that after its very first provisional formulation by Cowling (1934) researchers in the field tried to improve the theorem with respect to greater mathematical rigor, stronger statements or weaker assumptions. The history of this endeavour has already been reviewed in sufficient detail elsewhere. We mention only the excellent articles by Ivers \& James (1984), who cover the situation up to 1983,
by Fearn et al.\ (1988), who specify in particular some weak points and open problems, and by N$\acute{\rm u}\tilde{\rm n}$ez (1996), whose review presents the state of the art and emphasizes especially the role of Lortz and his coworkers. Proctor (2007) notes that this endeavour is not yet finished.
So, we refrain from giving another detailed account of the history but distinguish roughly three phases in the development of the axisymmetric theorem. 

In the first phase the {\em stationary} problem with solenoidal flow field and constant conductivity was in the focus. Cowling himself demonstrated the impossibility of a purely meridional magnetic field sustained by a purely meridional flow field by means of his now famous `neutral point argument'. Backus \& Chandrasekhar (1956) generalized this result by admitting meridional as well as azimuthal components for the magnetic field and the flow field, and they replaced the somewhat informal neutral point argument by mathematically more rigorous maximum principles for elliptic equations. Still later Lortz (1968) realized that these elliptic equations are incompatible with nontrivial solutions even for non-solenoidal flow fields and spatially variable conductivity. 

In the second phase the more realistic {\em time-dependent} problem came into focus but still under the proviso of a solenoidal flow field and constant conductivity. With these assumptions energy methods and variational inequalities can be applied to prove exponential decay in the energy norm of the meridional scalar and, in the absence of the meridional field, of the azimuthal field. Backus (1957) and Braginskii (1965) did so and gave, moreover, for balls explicit (albeit different) decay rate bounds on these quantities. Note that Backus (1957) considered already at that time the meridional problem without the solenoidal-flow and constant-conductivity asumptions and found exponential decay; however, in order to evaluate the variational problem, he had to introduce unsatisfactory ad-hoc assumptions.

The third phase was initiated by a paper of Todoeschuck \& Rochester (1980), who, in the view of Saturn's almost axisymmetric magnetic field, argued that Cowling's theorem might have loopholes, in particular, that non-stationary axisymmetric magnetic fields could be generated by non-solenoidal flows in a compressible fluid. Several researchers, however, dismissed this possibility and tried to close the loopholes; so, the third phase was characterized by attempts to prove the axisymmetric theorem for {\em non-solenoidal flows} and {\em variable conductivity}. The first such attempt was due to Hide \& and Palmer (1982), who demonstrated monotonous decay of the meridional scalar in the maximum norm. Their arguments were ingenious but, from a mathematical point of view, not rigorous (cf.\ 
N$\acute{\rm u}\tilde{\rm n}$ez 1996). Lortz \& Meyer-Spasche (1982a) gave a rigorous proof of this result using a clever combination of parabolic and elliptic maximum principles. Ivers \& James (1984) combining these maximum principles with skillfully chosen comparison functions strengthened this result in that they showed exponential decay to zero together with explicit decay rate bounds.

Less is known about the azimuthal field. First of all the azimuthal problem so far has exclusively been considered in the case that the meridional field has already died out and does not provide a source term for the azimuthal field. With this assumption Lortz \& Meyer-Spasche (1982b) proved the azimuthal field to be bounded in an integral norm for all times in terms of its initial value. Their method of proof depends on the existence of positive solutions of an auxiliary problem, which requires some mathematical effort. Ivers \& James (1984) obtained the same result using some more intuitive arguments. Finally, using Harnack-type inequalities for positive solutions of parabolic equations, Lortz, Meyer-Spasche \& Stredulinsky (1984) proved in a {\em plane} model problem exponential decay of the azimuthal field to zero even in the maximum norm.

Despite all these achievements there are still a number of shortcomings that need to be removed before the axisymmetric theorem can be regarded as fully established:

(i) In an axisymmetric setting the evolution of the meridional field decouples from that of the azimuthal field (which is the essence of the axisymmetric theorem), but not vice versa. Nevertheless, the evolution of the azimuthal field so far has exclusively been considered without meridional source term. So, a proof of the axisymmetric theorem using the full (coupled) set of dynamo equations is still missing.

(ii) Meridional decay results refer so far to the meridional scalar, for which a tractable equation is at hand, and not to the meridional field itself, with two notable exceptions: Hide (1981) argued that a signed version of the total magnetic flux leaving/entering the conductor has to decay in the case of axisymmetry; his arguments, however, are not rigorous. Ivers \& James (1984) appeal to Schauder estimates for elliptic and parabolic equations to relate the decay of the meridional field to that of the meridional scalar; these estimates, however, are not uniform in space; they fail, in particular, at the magnetic axis and at the conductor's boundary leaving the possibility - however unlikely - of magnetic regeneration there.

(iii) Even in the absence of the meridional source term, unconditional decay to zero of the azimuthal field is not yet established. Ivers \& James (1984) discuss a `scenario for decay to zero' based on additional ad-hoc assumptions, whereas the result of Lortz et al.\  (1984), although highly suggestive, does not make a statement about the {\em axisymmetric} problem. 

(iv) The only explicit decay rate bounds (from below) valid for general flow fields and conductivity distributions refer to the meridional field and are due to Ivers \& James (1984). These bounds become extremely small already for moderate values of the magnetic Reynolds number $R_m$. For instance, considering the earth, one finds for $R_m = 10^2$ (which is a rough estimate) magnetic decay rate bounds such that the corresponding decay times greatly exceed the lifetime of the earth. So, the question arises how close these bounds are to actual decay rates.

(v) In the case of solenoidal flow and constant conductivity lower decay rate bounds can be obtained by variational methods. Considering this variational problem for the meridional field in a ball Backus (1957) derived a bound strictly below the corresponding free decay rate whereas Braginskii (1965) noted (without proof) the free decay rate as (optimal) bound. Braginskii's claim, however, has been dismissed by
Dudley et al.\ (1986), who found by numerical computation the example of a meridional flow field that strictly slowed the free decay of the meridional field. A similar discrepancy in bounds given by Backus and Braginskii exists in the azimuthal problem with zero source term.  

We prove in the following unconditional, pointwise and exponential decay to zero of the meridional and the azimuthal fields satisfying the coupled dynamo equations, thus settling, finally, problems (i) to (iii) of the above list. Problem (iv) is resolved by example. We prove for a special piecewise constant flow field that the meridional decay rate shrinks exponentially fast to zero with respect to $R_m$. 
Numerical results for a smoothed version of this flow corroborate this outcome, which demonstrates that without further restrictions on the fluid flow no better decay rate bounds can be expected. Such restrictions can be formulated either geometrically for the velocity field or
in terms of the compressibility of the fluid. In fluids whose mass density and conductivity vary only weakly in a certain sense we show that the meridional and the azimuthal decay rates are - in orders of magnitude - comparable to those of free decay. 
Concerning the last problem (v) we corroborate in the meridional as well as in the azimuthal case Backus' lower bounds (by stricter and simpler arguments than those given by Backus) 
and dismiss Braginskii's claims by analytic counterexamples in the variational problems.

In proving these results we make use of all the mathematical tools already applied by previous authors, in particular, we use variational methods as in (Backus 1957), comparison functions as in (Ivers \& James 1984) and positive auxiliary solutions as in (Lortz et al.\ 1984). Most important, however, is a consistent use of the 5-dimensional formulation of the axisymmetric problem as initiated for the azimuthal field by Backus \& Chandrasekhar (1956).
This formulation eliminates the coordinate singularity inherent to the 3-dimensional axisymmetric formulation and makes the problem accessible to more or less standard techniques for partial differential equations. For instance, Galerkin representations of solutions are essential in the present context. These are standard for the azimuthal field and have recently been provided for the meridional scalar (Kaiser \& Uecker 2009). For solutions in this form we can use a `higher-order-decay theorem' relating the decay of the solution to the decay of its higher derivatives (Kaiser 2013).  This result allows, in particular, to relate the decay of the meridional scalar to that of the meridional field, and, moreover, to relate integral decay to pointwise decay. 

Concerning the organization of the material we have tried to make the paper accessible also to those readers who are less interested in all the mathematical details but are interested in the status of the axisymmetric theorem. So, some technical results are only cited 
and the pertinent calculations or proofs have either been omitted (and published elsewhere) or have been shifted to appendices, and the trustful reader might concentrate on the conclusions in the main text. In more detail the paper is organized as follows: section 2 presents the formulation of the axisymmetric dynamo problem in 2, 3 and 5 dimensions, which emphasize different aspects of the problem. In section 3 the higher-order-decay theorem is presented and then applied to the meridional decay problem whereas in section 4 results by Lortz et al.\ (1984) are adjusted to the axisymmetric situation and then applied to the azimuthal decay problem. A more technical part of this section has been shifted to appendix A. Section 5 is devoted to the construction of a meridional supersolution for a certain (discontinuous) flow field
to demonstrate the `slow' decay of the meridional field in this case; numerical results for a continuous version of this flow are also presented. Some more technical and some auxiliary material have again been deferred: appendix B contains the proof of a pertinent maximum principle and appendix C demonstrates the `fast' meridional decay for `one-dimensional' flows.
Section 6, finally, discusses lower meridional and azimuthal decay rate bounds obtained by variational methods which apply to incompressible fluids of constant conductivity. Here, an extension of these techniques is presented to the case of weakly varying density and conductivity.
\section{Formulation of the axisymmetric dynamo problem}
In the framework of magnetohydrodynamics the kinematic dynamo problem is the following initial value problem in all space (Backus 1958, Moffatt 1978) 
\be \label{2.1}
\le.
\ba{cl}
\pa_t \BB = -\na\ti (\eta \na\ti\BB) + \na\ti(\vv\ti\BB)\, ,\,\, \na\cdot\BB = 0 &\mbox{ in } G \ti \R_+ , \\[.5em]
\na\ti\BB = 0\, ,\quad \na\cdot\BB = 0 & \mbox{ in } \Gh \ti \R_+ , \\[.5em]
\BB \mbox{ continuous } & \mbox{ in } \R^3 \ti \R_+ , \\[.5em]
\BB(\xx,\cdot) \ra 0 & \mbox{ for } |\xx| \ra \infty, \\[.5em]
\BB(\cdot\, ,0) = \BB_0\, , \quad \na\cdot\BB_0 = 0 & \mbox{ on } G \ti \{t=0\}.
\ea
\ri\}
\ee
Here, the induction equation (\ref{2.1})$_1$ describes the generation of the magnetic field $\BB$ by the motion (with prescribed flow field $\vv$) of a conducting fluid (with magnetic diffusivity $\eta > 0$) in a bounded region $G \subset \R^3$. Outside the fluid region there are no further sources of magnetic field. Thus, $\BB$ matches continuously to some vacuum field in $\Gh := \R^3 \setminus \Go$ that vanishes at spatial infinity.

The central simplifying assumption of the present paper is axisymmetry of all variables appearing in (\ref{2.1}) including the shape $\pa G$ of the conductor. Using cylindrical coordinates $(\rho,\phi,z)$ with $\eee_z$ pointing in the direction of the symmetry axis $S$ and $(\rho, \phi)$ being polar coordinates in the planes perpendicular to $S$, this assumption implies the following representation of the solenoidal field $\BB$
\begin{equation}
\label{2.2}
\BB = \na M \ti \na \phi + A \na \phi = \Big(- \frac{1}{\rho}\, \pa_z M\, \eee_\rho + \frac{1}{\rho}\, \pa_\rho M\, \eee_z \Big) + \frac{1}{\rho}\, A\, \eee_\phi = \MM + \AA .
\end{equation}
The meridional field $\MM$ is confined to planes $\phi = const$ and thus everywhere perpendicular to the azimuthal field $\AA$. The corresponding scalars $M$ and $A$ depend on $\rho$, $z$ and $t$ but not on $\phi$.

Inserting (\ref{2.2}) into (\ref{2.1})$_{1a}$ and using the abbreviation $\EE := \eta \na \ti \BB - \vv \ti \BB$ yields componentwise
\be \label{2.3}
\le.
\ba{cl}
\pa_z(\pa_t M + \rho E_\phi ) = \pa_\phi E_z , \\[.5em]
\pa_t A + \pa_z E_\rho - \pa_\rho E_z = 0 , \\[.5em]
\pa_\rho (\pa_t M + \rho E_\phi ) = \pa_\phi E_\rho .
\ea
\ri\}
\ee
The right-hand sides of (\ref{2.3})$_{1,3}$ vanish by assumption and one concludes 
\be 
\label{2.4}
\pa_t M + \rho \big(\eta \na \ti (\na M \ti \na \phi + A \na \phi) - \vv \ti (\na M \ti \na \phi + A \na \phi)\big)_\phi = c(t) ,
\ee
where $c(t)$ is an undetermined function of $t$. Evaluating the left-hand side of (\ref{2.4}) and replacing $M- \int^t c(\tau)\, \rd \tau$ by $M$, which does not affect the meridional field, we arrive at
\be 
\label{2.5}
\pa_t M - \eta \Delta_* M + v_\rho\, \pa_\rho M + v_z\, \pa_z M = 0 ,
\ee
where $\D_*$ denotes the elliptic operator $\pa_\rho^2 - 1/\rho\, \pa_\rho + \pa_z^2$.

On the other hand, evaluating (\ref{2.3})$_2$ yields an evolution equation for the azimuthal scalar $A$ with a source term depending on $M$:
\be 
\label{2.6}
\pa_t A  - \rho\, \pa_\rho \Big(\frac{\eta}{\rho}\, \pa_\rho A\Big) - \pa_z (\eta\, \pa_z A) + \rho\, \pa_\rho \Big(\frac{v_\rho}{\rho}\, A\Big) + \pa_z (v_z\, \pa_z A) = - \rho\, \pa_\rho \Big(\frac{v_\phi}{\rho}\Big) \pa_z M + \pa_z v_\phi\, \pa_\rho M . 
\ee
Similarly but simpler, inserting (\ref{2.2}) into (\ref{2.1})$_{2a}$ yields in the vacuum region
\be
\label{2.7}
\D_* M = 0
\ee
and $\pa_\rho A = \pa_z A = 0$, which implies by the asymptotic condition (\ref{2.1})$_4$,
\be
\label{2.8}
A = 0 .
\ee
The matching condition (\ref{2.1})$_3$ implies for $M$ the matching condition
\be
\label{2.9}
M ,\; \pa_\rho M ,\; \pa_z M \;\; \mbox{continuous} 
\ee
and for $A$ by (\ref{2.8}) the boundary condition
\be
\label{2.10}
A\big|_{\pa G} = 0 .
\ee
The asymptotic condition (\ref{2.1})$_4$ reads for $M$
\be
\label{2.11}
\frac{1}{\rho}\, \pa_\rho M (\rho, z, \cdot) \ra 0\, ,\; \frac{1}{\rho}\, \pa_z M (\rho, z, \cdot) \ra 0\, \; \mbox{ for }\; \rho^2 + z^2 \ra \infty , 
\ee
where this notation implies that the convergence is uniform on compact sets of the variables represented by dots.
The coordinate singularity at $\rho = 0$ inherent to the representation (\ref{2.2}) requires additional conditions on $M$ and $A$ to avoid spurious solutions:
\be
\label{2.12}
M (\rho, \cdot, \cdot) \ra 0\, ,\; \pa_\rho M (\rho, \cdot, \cdot) = O(\rho)\, ,\; \pa_z M (\rho, \cdot, \cdot) = O(\rho)
\, \; \mbox{ for }\; \rho \ra 0
\ee
and
\be
\label{2.13}
A (\rho, \cdot, \cdot) = O(\rho^2)\, \; \mbox{ for }\; \rho \ra 0 . 
\ee
The conditions (\ref{2.12})$_{b,c}$ ensure a finite magnetic field on the symmetry axis $S$ and imply, moreover, the finite limit $\lim_{\rho \ra 0} M(\rho, \cdot,t) =: M_S (t)$. As $M_S(t)$ does not affect the meridional field we set it to zero. Condition (\ref{2.13}) ensures a differentiable (with respect to $\rho$) magnetic field component $B_\phi$ that vanishes at $S$. Finally, $\R^3$ is replaced by the half-plane $H := \{(\rho,z)\in \R_+ \ti \R\}$ and $G$ and $\Gh$ by the `cross-sections' $G_2 := \{(\rho,z)\in H: \xx(\rho, 0, z) \in G\}$ and $\Gh_2 := H\setminus \overline{G_2}$, respectively.

So, summarizing the foregoing results, a solution of the axisymmetric dynamo problem consists in a couple $(M,A)$ with $M$ satisfying (\ref{2.5}) in $G_2 \ti \R_+$ with initial value $M_0$, (\ref{2.7}) in $\Gh_2 \ti \R_+$ and (\ref{2.9}), (\ref{2.11}) and (\ref{2.12}) in $H \ti \R_+$ and, furthermore, with $A$ satisfying (\ref{2.6}) and (\ref{2.10}) in $G_2 \ti \R_+$ with initial value $A_0$. The initial values themselves have to satisfy (\ref{2.12}) and (\ref{2.13}) and typically some `compatibility conditions' (cf.\ Kaiser 2012).

This formulation in terms of the potentials $M$ and $A$ and the spatial variables $\rho$ and $z$ may be called minimal since it completely resolves the condition of axisymmetry. Most important, it yields an evolution equation for $M$ decoupled from $A$ and without zeroth-order term. However, there are also drawbacks. The divergence character of the azimuthal equation is veiled and, more important, we have to deal with singular coefficients in both equations. In fact, more redundant formulations turn out to be useful using cartesian coordinates in $\R^3$ and even $\R^5$ and using modified potentials. 

The 3-dimensional formulation is straightforward. With $\xx = (x,y,z) \in \R^3$ and the abbreviation $\rho := (x^2 +y^2)^{1/2}$ the meridional problem can be summarized as follows:\footnote{To keep the notation simple we do not use different symbols for the same function depending on different coordinates. The arguments should be clear from the context.}
\be \label{2.14}
\le.
\ba{cl}
\ds \pa_t M - \eta \D M + 2\eta\, \frac{\na \rho}{\rho} \cdot \na M + \vv \cdot \na M = 0 & \mbox{ in } G \ti \R_+ , \\[.8em]
\ds \D M - 2\, \frac{\na \rho}{\rho} \cdot \na M = 0 & \mbox{ in } \Gh \ti \R_+ , \\[.8em]
M \mbox{ and } \na M \mbox{ continuous } & \mbox{ in } \R^3 \ti \R_+ , \\[.5em]
M(x, y, \cdot, \cdot ) \ra 0\; ,\,\na M(x, y, \cdot, \cdot) = O(\rho) & \mbox{ for } \rho \ra 0, \\[.5em]
\ds \frac{1}{\rho}\,\na M(\xx,\cdot) \ra 0 & \mbox{ for } |\xx| \ra \infty, \\[.8em]
M(\cdot\, ,0) = M_0 & \mbox{ on } G \ti \{t=0\}.
\ea
\ri\}
\ee
Here, $\na$ denotes the Cartesian gradient vector $(\pa_x, \pa_y,\pa_z)$ and in (\ref{2.14})$_{1,2}$ we made use of the identity for axisymmetric functions,
$$
\D_* M = \pa_\rho^2 M - \frac{1}{\rho}\, \pa_\rho M + \pa_z^2 M = \D M - 2\, \frac{\na \rho}{\rho} \cdot \na M .
$$

Concerning the azimuthal problem we introduce the variable $\cA := A/\rho^2$ and use the identity for axisymmetric functions
$$
\frac{1}{\rho}\, \pa_\rho \Big(\frac{\eta}{\rho}\, \pa_\rho \big(\rho^2 \cA\big)\Big) +\pa_z \big( \eta\, \pa_z \cA\big) = \na \cdot \Big( \frac{\eta}{\rho^2} \na\big(\rho^2 \cA\big)\Big)
$$
to obtain
\be \label{2.15}
\le.
\ba{cl}
\ds \pa_t \cA - \na \cdot \Big( \frac{\eta}{\rho^2} \na\big(\rho^2 \cA\big)\Big) +\na \cdot (\vv \cA) = \na \cdot \Big(\MM\, \frac{v_\phi}{\rho}\Big) & \mbox{ in } G \ti \R_+ , \\[.5em]
\cA = 0 & \mbox{ on } \pa G \ti \R_+ , \\[.5em]
\cA(\cdot\, ,0) = \cA_0 & \mbox{ on } G \ti \{t=0\}.
\ea
\ri\}
\ee
Note that a solution of (\ref{2.15}) that is well-defined on the symmetry axis automatically satisfies (\ref{2.13}).

The 3-dimensional formulation of the azimuthal problem makes the divergence character manifest; the coefficients, however, are still singular. This problem can be cured by a five-dimensional formulation. Let $\xx = (x_1, x_2, x_3, x_4, x_5) \in \R^5$, $\na$ the corresponding cartesian gradient  and $\D$ the 5-dimensional Laplacian. Identifying $\rho^2$ with $\sum_{i=1}^4 x_i^2$ and $z$ with $x_5$ and defining $G_5 := \{\xx \in \R^5 : (\rho, z) \in G_2\}$, axisymmetric functions in $\R^3$ can be considered as axisymmetric functions in $\R^5$. Introducing the variable $\cM := M/\rho^2$ and the 5-dimensional meridional flow field $\vv_m^{(5)} := \sum_{i=1}^4 (v_\rho/\rho) x_i \eee_i + v_z \eee_5$, and observing the axisymmetric identity
$$
\D_* M = \frac{1}{\rho}\, \pa_\rho \Big( \rho^3 \pa_\rho \Big(\frac{1}{\rho^2}\, M\Big)\Big) + \pa_z^2 M = \rho^2 \na \cdot \na \Big(\frac{1}{\rho^2}\, M \Big) = \rho^2 \D \cM ,
$$
the meridional problem takes the form
\be \label{2.16}
\le.
\ba{cl}
\ds \pa_t \cM - \eta \D \cM + \vv_m^{(5)} \cdot \na \cM + 2\,\frac{v_\rho}{\rho}\, \cM = 0 & \mbox{ in } G_5 \ti \R_+ , \\[.5em]
\D \cM = 0 & \mbox{ in } \Gh_5 \ti \R_+ , \\[.5em]
\cM \mbox{ and } \na \cM \mbox{ continuous } & \mbox{ in } \R^5 \ti \R_+ , \\[.5em]
\cM(\xx,\cdot) \ra 0 & \mbox{ for } |\xx| \ra \infty, \\[.5em]
\cM(\cdot\, ,0) = \cM_0 & \mbox{ on } G_5 \ti \{t=0\}.
\ea
\ri\}
\ee
Note that for a smooth axisymmetric flow field $v_\rho/\rho$ is well-defined on $S$; thus, no singular coefficients appear any more in (\ref{2.16}). No axis-condition on $\cM$ is anymore necessary; condition (\ref{2.12}) is automatically satisfied for any well-defined solution of (\ref{2.16}). Moreover, outside the conductor, $\cM$ is a harmonic potential (in 5 dimensions); using this information (cf.\  Kaiser \& Uecker 2009, Appendix C) condition (\ref{2.16})$_4$ implies, in particular,
$$
\na \cM(\xx,\cdot) = O(|\xx|^{-4}) \quad \mbox{ for } |\xx| \ra \infty , 
$$
which again implies (\ref{2.11}). Note, however, that for these advantages we had to pay a price, viz.\  a zeroth-order term in (\ref{2.16})$_1$.

Concerning the azimuthal problem we make use of the axisymmetric identities 
$$
\rho\, \pa_\rho \Big(\frac{\eta}{\rho}\, \pa_\rho A\Big) +\pa_z \big( \eta\, \pa_z A\big) = \rho^2\, \na \cdot \Big( \eta \na \frac{A}{\rho^2} \Big) + 2\, \frac{\pa_\rho \eta}{\rho}\, A
$$
and 
$$
\rho\, \pa_\rho \Big(\frac{v_\rho}{\rho}\, A\Big) +\pa_z v_z\, A = \rho^2\, \na \cdot \Big( \vv_m^{(5)}\, \frac{A}{\rho^2} \Big) - 2\, \frac{v_\rho}{\rho}\, A
$$
to obtain
\be \label{2.17}
\le.
\ba{cl}
\ds \pa_t \cA - \na \cdot (\eta \na \cA) +\na \cdot (\vv_m^{(5)} \cA) - 2 \Big(\frac{\pa_\rho \eta}{\rho} + \frac{v_\rho}{\rho}\Big) \cA & \\[.5em]
\ds = \rd\vv_a^{(5)} \cdot \na \cM + 2\, \frac{\pa_z v_\phi}{\rho}\, \cM & \mbox{ in } G_5 \ti \R_+ , \\[.5em]
\cA = 0 & \mbox{ on } \pa G_5 \ti \R_+ , \\[.5em]
\cA(\cdot\, ,0) = \cA_0 & \mbox{ on } G_5 \ti \{t=0\} ,
\ea
\ri\}
\ee
where we used the abbreviation $\rd\vv_a^{(5)} := \sum_{i=0}^4 \pa_z v_\phi\, (x_i /\rho) \eee_i - \rho\, \pa_\rho (v_\phi /\rho) \eee_5$. 
On the premises of smooth (up to second order) axisymmetric data $\eta$ and $\vv$, all coefficients in (\ref{2.17}) are now well-defined on $S$ (cf.\ Ivers \& James 1984). But, again, we have traded for this advantage a zeroth-order term that destroys the divergence character of the left-hand side in (\ref{2.17})$_1$. 

As far as the various forms of meridional and azimuthal evolution equations have been considered by previous authors, these coincide, of course, with ours. Concerning the asymptotic conditions at $S$ and at infinity, however, there are some minor differences. For example, Ivers \& James (1984) use the condition
\be
\label{2.18}
M (\rho, z, \cdot) = O\big((\rho^2 + z^2)^{-1/2}\big)\, \; \mbox{ for }\; \rho^2 + z^2 \ra \infty  
\ee
instead of (\ref{2.11}). Stredulinski et al.\ (1986) called condition (\ref{2.18}) in question. The discrepancy, however, seems to be due to a confusion about the axisymmetric and the plane-symmetric problem (cf.\ Kaiser \& Uecker 2009, Remark B.1). In fact, in view of the five-dimensional formulation conditions (\ref{2.11}) and (\ref{2.18}) are equivalent for solutions of (\ref{2.7}). A similar remark holds for the axis-condition (\ref{2.12}), which is typically replaced by previous authors (as far as they specify an axis-condition at all) by the slightly weaker condition  
\be
\label{2.19}
M (\rho, \cdot , \cdot) = O(\rho^2)\, \; \mbox{ for }\; \rho \ra 0 .
\ee
In fact, (\ref{2.19}) is enough to establish the equivalence between the two- and the five-dimensional problem and hence the equivalence between (\ref{2.12}) and (\ref{2.19}) for meridional solutions.

Concerning the flow field some authors require the normal component to vanish on $\pa G$ whereas other authors do not mention any boundary conditions. In fact, some results depend on such a condition, others do not. We will specify in the following this condition where needed.

A final remark concerns the question, to which quantities  axisymmetry has exactly to apply. In view of eqs.\ (\ref{2.3}) Lortz (1968) required axisymmetry of the electrical quantities $\BB$ and $\EE$ but not of $\vv$ or $\eta$. Todoeschuck \& Rochester (1980) critized this approach as a disguise of the usual requirements.  Lortz's assumptions are indeed enough to derive eqs.\ (\ref{2.5}) and (\ref{2.6}) even for non-axisymmetric coefficients $\vv$ and $\eta$. However, evaluating the conditions $\pa_\phi E_\rho = \pa_\phi E_z = 0$ on their part yield the additional equations 
\be \label{2.20}
\le.
\ba{ll}
\pa_\phi \eta\, \pa_\rho A - \pa_\phi v_\rho\, A - \pa_\phi v_\phi\, \pa_z M = 0 , \\[.5em]
\pa_\phi \eta\, \pa_z A - \pa_\phi v_z\, A + \pa_\phi v_\phi\, \pa_\rho M = 0 . 
\ea
\ri\}
\ee
These equations were trivially solved by axisymmetric coefficients (and this solution Todoeschuck \& Rochester probably had in mind). Otherwise, for prescribed non-axisymmetric coefficients, (\ref{2.20}) represent additional constraints. The combined system (\ref{2.5}), 
(\ref{2.6}) and (\ref{2.20}) for axisymmetric scalars $M$ and $A$, however, does not seem to be well-posed; in particular, general solvability cannot be expected. So, the present treatment of the axisymmetric problem cannot dispense with the assumption of axisymmetric coefficients. 
\section{Decay of meridional field and azimuthal current}
In this section the exponential decay of the meridional scalar $M$ in the maximum norm is related to exponential decay with (almost) the same decay rate of the meridional field $\MM$ and of the azimuthal current $\JJ_a$. As discussed in the introduction pointwise decay of $M$, for instance in the form 
\be \label{3.1}
\ds |M(\rho, \cdot, t) | \leq g(\rho)\, e^{- d_0 t} ,\qquad \rho >0\, ,\; t \geq 0
\ee
with some $d_0 >0$ and a function $0 < g(\rho) < C$ with (at least) $g(\rho) = O(\rho)$ for $\rho \ra 0$, is well established (see Ivers \& James 1984) and is the prerequisite of the present section. We come back to this subject when discussing decay rates in sections 5 and 6.

The central tool of this section is the following higher-order-decay theorem for smooth solutions $u$ of the system 
\be \label{3.2}
\le.
\ba{cl}
\ds \pa_t u - a \D u + \bb \cdot \na u + c\, u = f & \mbox{ in } G \ti \R_+ , \\[.5em]
\D u = 0 & \mbox{ in } \Gh \ti \R_+ , \\[.5em]
u \mbox{ and } \na u \mbox{ continuous } & \mbox{ in } \R^n \ti \R_+ , \\[.5em]
u(\xx,\cdot) \ra 0 & \mbox{ for } |\xx| \ra \infty, \\[.5em]
u(\cdot\, ,0) = u_0 & \mbox{ on } G \ti \{t=0\}.
\ea
\ri\}
\ee
Here, $G$ is a bounded domain in $\R^n$ $(n \geq 3)$ with complement $\Gh = \R^n \setminus \Go$ and (sufficiently smooth) boundary $\pa G$. Associated to (\ref{3.2}) is the eigenvalue problem  
\be \label{3.3}
\le.
\ba{cl}
-\D u = \la u & \mbox{ in } G , \\[.5em]
\D u = 0 & \mbox{ in } \Gh , \\[.5em]
u \mbox{ and } \na u \mbox{ continuous } & \mbox{ in } \R^n , \\[.5em]
u(\xx) \ra 0 & \mbox{ for } |\xx| \ra \infty, 
\ea
\ri\}
\ee
which is well-defined and has a lowest eigenvalue $\la_1$ that is positive (Kaiser \& Uecker 2009).

To measure the spatial smoothness of the coefficients $a$, $\bb$ and $c$ we use the notation $a \in C^k (\Go \ti \R_+)$ $(k\geq 0)$,
which means that all spatial derivatives up to order $k$ are continuous and satisfy bounds of the form\footnote{Here, $\al$ enumerates all derivatives $D^\al$ of order $\leq k$.} $|D^k a|_{max} := \max_{|\al| \leq k} \sup_{\Go \ti \R_+} |D^\al a| < K$ for some $K > 0$. $a \in C_1^k (\Go \ti \R_+)$ means that, additionally, $\pa_t u$ is a continuous function with corresponding bound.
\begin{Theorem}[Higher-order-decay]
Let $a,\, \bb,\, c \in C^k (\Go \ti \R_+)$ $(k\geq 0, even)$ with common bound $|D^k a|_{max},\, |D^k \bb|_{max},\, |D^k c|_{max} < K$ and $a\geq a_0 > 0$. Let, furthermore, $f$ satisfy the higher-order energy decay condition
\be \label{3.4}
\sum_{|\al| \leq k} \int_G|D^\al f(\xx, t)|^2 \rd x \leq C_f\, e^{-2 d_f t} , \qquad t\geq 0
\ee
with some constants $C_f >0$ and $d_f >0$, and let $u$ be a classical (i.e.\ $u \in C_1^2 (G \ti \R_+))$ solution of (\ref{3.2}). Then, the energy decay condition 
\be \label{3.5}
\int_G|u(\xx, t)|^2 \rd x \leq C_0\, e^{-2 d_0 t} , \qquad t \geq 0
\ee
implies the higher-order energy decay
\be \label{3.6}
\sum_{|\al| \leq k+1} \int_G|D^\al u(\xx, t)|^2 \rd x \leq C\, e^{-2 d t} , \qquad t\geq 0
\ee
as well as decay in the maximum norm:
\be \label{3.7}
\sum_{|\al| = l} \sup_{\Go} |D^\al u(\cdot, t)| \leq \Cti\, e^{-d t} , \qquad t\geq 0 ,
\ee
where $l < k+1-n/2$. Here, $d:= \min \{d_1 - \ep, d_0, d_f\}$, where $d_1 := a_0 \la_1$ is the `free' decay rate of (\ref{3.2}), and $C$ and $\Cti$ are constants depending on $G$, the initial value $u_0$, $K$, $C_0$, $C_f$,$n$, $k$ and $\ep >0$.
\end{Theorem}
The proof of this rather technical result can be found in (Kaiser 2013). Yet some remarks are in order: the existence of classical solutions of the required type depends on the regularity of the coefficients (as specified in the theorem), on the regularity of the initial value $u_0$ and on suitable `compatibility conditions' (for details see (Kaiser \& Uecker 2009)). So, as a rule, for sufficiently regular data, conditions (\ref{3.4}) and (\ref{3.5}) imply decay to arbitrarily large orders. The decay rate of $u$ is bounded by the lowest of the three decay rates $d_1$, $d_0$ and $d_f$; the free one $d_1$, however, can only `almost' be attained since typically $C$ and $\Cti$ diverge (at most algebraically) in the limit $\ep \ra 0$.

The 5-dimensional formulation (\ref{2.16}) of the meridional problem is obviously of the type to which theorem 1 can be applied. So, setting $f=0$, $n =5$ and $k=4$ the remaining condition (\ref{3.5}) takes the form
\be \label{3.8}
\int_{G_5} \cM(\xx, t)^2 \rd x = |S_3| \int_{G_2} \Big(\frac{1}{\rho^2} M(\rho,z,t)\Big)^2 \rho^3 \rd \rho\, \rd z \leq C_0\, e^{-2 d_0 t} ,
\ee
where $|S_3|$ is the volume of the 3-dimensional sphere $S_3$. Condition (\ref{3.8}) is clearly implied by (\ref{3.1}) when observing the axis-condition of the radial function $g(\rho)$.

As consequences we have then (\ref{3.6}):
\be \label{3.9}
\sum_{|\al| \leq 5} \int_{G_5} |D^\al \cM(\xx, t)|^2 \rd x \leq C\, e^{-2 d t} , \qquad t\geq 0
\ee
and (\ref{3.7}) with $l = 0, 1$ and $2$:
\be \label{3.10}
\sup_{\Go_5} |\cM(\cdot, t)| + \sum_{i=1}^5 \sup_{\Go_5} |\pa_{x_i} \cM(\cdot, t)| \leq \Cti\, e^{-d t} , \qquad t\geq 0 ,
\ee
\be \label{3.11}
\sum_{i,j =1}^5 \sup_{\Go_5} |\pa_{x_i} \pa_{x_j} \cM(\cdot, t)| \leq \Cti\, e^{-d t} , \qquad t\geq 0 ,
\ee
which yield uniform pointwise bounds on 
$$
|\MM| = \Big|\! -\frac{1}{\rho}\, \pa_z M \eee_\rho + \frac{1}{\rho}\, \pa_\rho M \eee_z \Big| = \frac {1}{\rho}\, |\na (\rho^2 \cM)| = |2\, \cM \na \rho + \rho \na \cM|
$$
and 
$$
|\JJ_a| = |\na \ti \MM| = \frac{1}{\rho}\, |\D_* M | = \rho |\D \cM| ,
$$
respectively. As $\cM$ is harmonic in $\Gh_5$, continuous in $\R^5$ and vanishes at infinity, $\cM$ takes its maximum and minimum in $\Go_5$.
This conclusion holds likewise for each component of $|\xx|\na \cM$. In fact, $|\xx|\pa_{x_i} \cM$ solves the elliptic equation with negative zeroth order term,
$$
\Big( \D - 2\, \frac{\xx}{|\xx|^2}\cdot \na - \frac{2}{|\xx|^2} \Big) u = 0 \quad \mbox{ in } \; \Gh_5 ,
$$
and, thus, obeys a maximum principle, too. With (\ref{3.10}) we can, therefore, estimate for $t \geq 0$:
\be \label{3.12}
\ba{rl}
\ds \sup_H |\MM (\cdot, \cdot,t)| \!\!\!
& \ds \leq 2 \sup_{\R^5} |\cM(\cdot, t)| + \sum_{i=1}^5 \sup_{\R^5} |\xx| |\pa_{x_i} \cM(\cdot, t)| \\[.7em]
& \ds \leq 2 \sup_{\Go_5} |\cM(\cdot, t)| + \sum_{i=1}^5 \sup_{\Go_5} |\xx| |\pa_{x_i} \cM(\cdot, t)| \\[1.2em]
& \ds \leq \max \{2,R\}\, \Cti\, e^{-d t} \, =: \Cti_{\MM}\, e^{-d_m t} ,
\ea
\ee
where $R$ is the radius of a ball enclosing the conductor. As $\JJ_a$ is restricted to the conductor one obtains directly by (\ref{3.11}):
\be \label{3.13} 
\sup_{\Go_2} \Big| \frac{1}{\rho}\, \JJ_a (\cdot,\cdot, t)\Big| \leq \sup_{\Go_5} |\D \cM (\cdot, t)| \leq \Cti_{\JJ_a}\, e^{- d_m t} ,\qquad t \geq 0 .
\ee
According to theorem 1 the decay rate bound $d_m$ is given by $d_m = \min \{d_1 - \ep , d_0 \}$. 
The bound (\ref{3.1}) holds for all flow fields and all resistivity distributions $\eta \geq \eta_0$; it covers, in particular, the free case; thus, $d_0 \leq d_1$. Comparing the decay of meridional field and meridional scalar we find, therefore, in the case $d_0 < d_1$ the same bound for both quantities and in the case $d_0 = d_1$ almost the same bound. Concerning the constants $\Cti_{\MM}$ and $\Cti_{\JJ_a}$ we made no effort to determine the exact dependence on the various parameters listed in theorem 1. So, numerical values could be very large. From an observational point of view, however, these can always be compensated by sufficiently large (compared to the decay time $1/d_m$) periods of time. Finally, the high degree of regularity of the data (boundedness of fourth-order derivatives) required by theorem 1 is due to the method of proof, in particular, the use of embedding results in $\R^5$ neglecting the axisymmetry of the problem.  Using more refined proof techniques the regularity requirements are expected to decrease.
\section{Decay of azimuthal field and meridional current}
In this section the unconditional decay of the azimuthal field $\AA$ and of the meridional current $\JJ_m$ is proved exploiting the divergence-character of the governing equation in 3 dimensions as well as the nonsingular formulation in 5 dimensions. A central tool is a positive axisymmetric solution $P$ of the following auxiliary problem  
\be \label{4.1}
\le.
\ba{cl}
\ds \pa_t P - \na \cdot \Big( \frac{\eta}{\rho^2} \na\big(\rho^2 P\big)\Big) +\na \cdot (\vv P) = 0 & \mbox{ in } G \ti \R_+ , \\[.5em]
\nn \cdot \na P = 0 & \mbox{ on } \pa G \ti \R_+ , \\[.5em]
P(\cdot\, ,0) = P_0  > 0 & \mbox{ on } G \ti \{t=0\},
\ea
\ri\}
\ee
that differs from (\ref{2.15}) in a vanishing right-hand side in (\ref{4.1})$_1$ and a Neumann boundary condition instead of a Dirichlet condition. All data $G$, $\eta$, $\vv$ and $P_0$ are assumed to be axisymmetric, which implies that the solution is axisymmetric, too. The regularity requirements that ensure the existence of solutions are moderate; they are more detailed for the non-singular 5-dimensional formulation in appendix A. We mention here only that $G \in \R^3$ is a bounded domain with smooth boundary $\pa G$ and exterior normal $\nn$, $\eta \in C^1(\Go \ti \R_+)$ and $\vv \in C (\Go \ti \R_+)$ with $\nn \cdot \vv|_{\pa G} = 0$. Moreover, we need the bounds $|\vv| \leq K$, $|\na \rho \cdot \vv/\rho| \leq K$, $|\na \rho \cdot \na\eta /\rho| \leq K$ for some $K>0$ and $\eta \geq \eta_0 > 0$ on $\Go \ti \R_+$.
\begin{Theorem}[Positive auxiliary solution]
Let $P$ be a solution of the axisymmetric problem (\ref{4.1}) with initial value $P_0$ and let $\Pu_0$, $\Po_0$ be some positive numbers that bound $P_0$:
\be \label{4.2}
\Pu_0 \leq P_0 \leq \Po_0 \qquad \mbox{ in } G .
\ee
Then, positive bounds $\Pu$ and $\Po$ exist such that  
\be \label{4.3}
\Pu \leq P(\cdot, t) \leq \Po \qquad \mbox{ for } t \geq 0 ,
\ee
where $\Pu$ and $\Po$ depend only on $G$, $\eta_0$, $K$ and $\Pu_0$ and $\Po_0$.
\end{Theorem}
The proof of theorem 2 consists in an adaptation to our situation of a powerful result by Lortz et al.\ (1984). The necessary changes are rather technical and are, therefore, deferred to appendix A.

We compute next the time-derivative of the quantity $\int_G \cA^2 /P \,\rd x$. To avoid singular coefficients we consider first a `regularized' domain $G_\ep := \{ \xx \in G: \rho(\xx) > \ep \}\; (\ep > 0)$ bounded by the surface $S_\ep := \{ \xx \in \pa G: \rho(\xx) > \ep \}$ and the cylinder $C_\ep := \{ \xx \in G: \rho(\xx) = \ep \}$ such that $\pa G_\ep = S_\ep \cup \Co_\ep$. In the limit $\ep \ra 0$, $C_\ep$ shrinks to the line segment $L := \{ \xx \in G: \rho(\xx) = 0 \}$. By (\ref{2.15})$_1$, (\ref{4.1})$_1$ and with the abbreviation $f := \MM \cdot \na (v_\phi / \rho)$ one obtains\footnote{For simplicity, dependence on $t$ is suppressed in the subsequent calculations.}
$$
\ba{c}
\ds \frac{\rd}{\rd t} \int_{G_\ep} \frac{\cA^2}{P}\, \rd x = \int_{G_\ep} \Big(2\, \frac{\cA}{P}\, \pa_t \cA - \frac{\cA^2}{P^2}\, \pa_t P\Big) \rd x\\[1em]
\ds =  2 \int_{G_\ep} \frac{\cA}{P}\, f\, \rd x +
\int_{G_\ep} \bigg\{2\, \frac{\cA}{P}\, \na \cdot \Big( \frac{\eta}{\rho^2} \na\big(\rho^2 \cA\big) -\vv \cA\Big)
-\frac{\cA^2}{P^2}\, \na \cdot \Big( \frac{\eta}{\rho^2} \na\big(\rho^2 P\big) - \vv P \Big)\bigg\} \rd x .
\ea
$$
Integrating by parts in the last integral and using the boundary condition (\ref{2.15})$_2$ on $S_\ep$, the velocity field $\vv$ drops from the integral and we are left with
$$
\ba{l}
\ds - \int_{G_\ep} \bigg\{2\, \na \Big(\frac{\cA}{P}\Big) \cdot \frac{\eta}{\rho^2} \na\big(\rho^2 \cA\big) 
-2\, \frac{\cA}{P}\, \na \Big(\frac{\cA}{P}\Big) \cdot \frac{\eta}{\rho^2} \na\big(\rho^2 P\big) \bigg\} \rd x \\[1em]
\ds - \int_{C_\ep} \bigg\{2\, \frac{\cA}{P}\, \frac{\eta}{\rho^2}\, \pa_\rho \big(\rho^2 \cA\big) - 
\frac{\cA^2}{P^2}\, \frac{\eta}{\rho^2}\, \pa_\rho \big(\rho^2 P\big) \bigg\} \rd s \\[1em]
\ds = - 2 \int_{G_\ep} \eta P \Big|\na \Big(\frac{\cA}{P}\Big)\Big|^2 \rd x - \int_{C_\ep} \eta\, \Big( \frac{\cA}{P}\, \pa_\rho \cA - \frac{\cA^2}{P^2}\, \pa_\rho P\Big) \rd s - 2 \int_{C_\ep} \frac{\eta}{\rho}\, \frac{\cA^2}{P}\, \rd s . 
\ea
$$
In the limit $\ep \ra 0$ only the second surface integral survives with the result $- 4\pi  \int_{L} \eta\, \cA^2/P\, \rd z$. In summary we have in the limit $\ep \ra 0$:
\be \label{4.4}
\ds \frac{\rd}{\rd t} \int_{G} \frac{\cA^2}{P}\, \rd x = 
- 2 \int_{G} \eta P \Big|\na \Big(\frac{\cA}{P}\Big)\Big|^2 \rd x - 4\pi  \int_{L} \eta\, \frac{\cA^2}{P}\, \rd z + 2 \int_{G} \frac{\cA}{P}\, f\, \rd x .
\ee
By means of the bounds (\ref{4.3}) on $P$, the variational inequality 
$$
\int_G |\na g|^2\, \rd x + 2\pi \int_L |g|^2 \,\rd z \geq \mu_1 \int_G |g|^2 \,\rd x
$$
for axisymmetric differentiable functions $g$ vanishing on $\pa G$ (see section 6), the bound (\ref{3.12}) on $\MM$ and the bound $|\na (v_\phi/\rho)| \leq K$, (\ref{4.4}) can be estimated as follows:
\be \label{4.5}
\ba{rl}
\ds \frac{\rd}{\rd t} \int_{G} \frac{\cA^2}{P}\, \rd x \!\!\!
&\ds\leq - 2 \eta_0\, \Pu \bigg\{ \int_{G} \Big|\na \Big(\frac{\cA}{P}\Big)\Big|^2 \rd x + 2\pi  \int_{L} \frac{\cA^2}{P^2}\, \rd z \bigg\} + 2 \int_{G} \frac{\cA}{P}\, \MM \cdot \na \Big(\frac{v_\phi}{\rho}\Big) \rd x \\[1em]
&\ds \leq - 2 \eta_0 \mu_1\, \Pu \int_{G} \frac{\cA^2}{P^2}\, \rd x + 2 K \Cti_{\MM}\, e^{- d_m t}\, \frac{1}{\Pu^{1/2}} \int_{G} \frac{\cA}{P^{1/2}}\, \rd x \\[1em]
&\ds \leq - 2 \eta_0 \mu_1 \frac{\Pu}{\Po} \int_{G} \frac{\cA^2}{P}\, \rd x + 2 K \Cti_{\MM}\, e^{- d_m t}\, \frac{|G|^{1/2}}{\Pu^{1/2}} \bigg(\int_{G} \frac{\cA^2}{P}\, \rd x\bigg)^{1/2} \\[1em]
&\ds \leq - 2 (d_a - \ep) \int_{G} \frac{\cA^2}{P}\, \rd x + \frac{K^2 \Cti_{\MM}^2 |G|}{2 \ep\, \Pu} \, e^{- 2 d_m t} .
\ea
\ee
In the third line we used Youngs's inequality to split the second term and introduced the azimuthal decay rate $d_a := \eta_0 \mu_1\,\Pu\, /\Po$. Applying Gronwall's inequality  on (\ref{4.5}) yields then the exponential decay of the quantity $\int_G \cA^2 /P\, \rd x$:
$$
\int_G \frac{\cA(\xx, t)}{P(\xx,t)}\, \rd x \leq \int_{G} \frac{\cA_0^2 (\xx)}{P_0 (\xx)}\, \rd x \,\, e^{-2(d_a-\ep) t} + \frac{K^2 \Cti_{\MM}^2 |G|}{2 \ep\, \Pu} \, \frac{e^{-2(d_a -\ep) t} - e^{- 2 d_m t}}{2(d_m -d_a + \ep)}\, ,\quad t\geq 0 .
$$
Eliminating, finally, the auxiliary function $P$, we end up with the energy estimate
\be \label{4.6}
\ba{rl}
\ds \int_G \cA^2 (\xx,t)\, \rd x \leq \!\!\!\! &\ds \Po \int_G \frac{\cA^2(\xx, t)}{P(\xx,t)}\, \rd x \leq \frac{\Po}{\Pu_0} \int_{G} \cA_0^2 (\xx)\, \rd x \,\, e^{-2(d_a-\ep) t} \\[1.5em] 
& \ds + \frac{\Po}{\Pu}\,\frac{K^2 \Cti_{\MM}^2 |G|}{4 \ep |d_m -d_a + \ep|} \,\, e^{-2 \min \{d_a -\ep,\, d_m\} t} =: C_A\, e^{-2 \min \{d_a -\ep,\, d_m\} t} ,
\ea
\ee
where $C_A$ depends on the initial value $\cA_0$, the fraction $\Po/\Pu$, $K$, $\Cti_{\MM}$, $|G|$, $d_m - d_a$ and $\ep > 0$ (chosen such that $d_m - d_a + \ep \neq 0$).

To obtain bounds on $\AA$ and $\JJ_m = \na \ti \AA$ in the maximum norm we use once more theorem 1, which holds literally (and even without restriction to even $k$)
also in the case that the `boundary condition' (\ref{3.2})$_{2,3,4}$ is replaced by a Dirichlet condition (Kaiser 2013). Resolving the divergence form, the governing equation (\ref{2.17})$_1$ of the azimuthal problem in $\R^5$ takes the form of (\ref{3.2})$_1$. So, theorem 1 may be applied to solutions of (\ref{2.17}). Choosing $n=5$ and $k=3$, the main conditions (\ref{3.4}) and (\ref{3.5}) are satisfied by (\ref{3.9}) with $d_f = d = d_m$ and (\ref{4.6}) with $d_0 = \min \{d_a -\ep,\, d_m\}$, respectively. Note that $d_a$ bounds, in particular, the free decay in (\ref{2.17}), thus $d_a \leq d_1$. By (\ref{3.10}) we obtain then the bound
$$
\sup_{\Go_5} |\cA(\cdot, t)| + \sup_{\Go_5} |\na \cA(\cdot, t)| \leq \Cti\, e^{-\min \{d_a - \ep, \, d_m\} t} , \qquad t\geq 0 
$$
and hence for all time
\be \label{4.7}
\sup_{\Go} \Big|\frac{1}{\rho}\,\AA(\cdot, t)\Big| = \sup_{\Go_5} |\cA(\cdot, t)| \leq \Cti_{\AA}\, e^{-\min \{d_a - \ep, \, d_m\} t} 
\ee
and 
\be \label{4.8}
\sup_{\Go} |\JJ_m (\cdot, t)| \leq 2 \sup_{\Go_5} |\cA(\cdot, t)| + R \sup_{\Go_5} |\na \cA(\cdot, t)| \leq \Cti_{\JJ_m}\, e^{-\min \{d_a - \ep, \, d_m\} t} . 
\ee
The constants $\Cti_{\AA}$ and $\Cti_{\JJ_m}$ depend, as in the meridional case, on bounds on the derivatives of the data $\eta$ and $\vv$ up to fourth order; moreover on $\eta_0$, the conducting region $G$, the initial values $\cM_0$ and $\cA_0$, and the `balance parameter'
$\ep$. All other dependencies can be eliminated by substitution; for instance the `oscillation' $\Po /\Pu$ of the auxiliary function $P$ depends, according to theorem 1, on no other parameters when the initial value $P_0 = const = 1$ is chosen.
\section{An example of slow decay}
In this section we present the example of a flow field, for which the meridional decay rate shrinks exponentially fast to zero with respect to the amplitude of the flow. The flow field has been chosen as simple as possible, viz.\ piecewise constant, in order to allow the explicit construction of a subsolution providing an upper bound on the meridional decay rate of any axisymmetric solution of the dynamo problem. The numerical investigation of a smoothed version of this flow field corroborates our analytic findings in the range of attainable flow amplitudes. Of course, this model flow is not likely to be a reasonable astrophysical flow but it makes clear that without further assumptions no better decay results can be expected than those of Ivers \& James (1984).  

In order to relate solutions with subsolutions of the dynamo problem we make use of the following maximum principle that is tailored to our needs: let $G \subset \R^n$ be a bounded domain with smooth boundary $\pa G$ and complement $\Gh$. Let, furthermore, $G$ be decomposed into  a finite number of subdomains $G_1, \ldots, G_k$ separated by smooth hypersurfaces $\Ga_1, \ldots, \Ga_l$ such that $G = \bigcup_{i=1}^k \Go_i$ and $G_i \cap G_j = \emptyset$ for $i \neq j$. So, $\pa G_i$ may have parts in $\Ga:= \bigcup_{j=1}^l \Ga_j$ and $\pa G$; it is smooth up to possible intersections of the bounding components $\Ga_j$ and $\pa G$. To take care of a possible symmetry axis $S$ in the $x_n$-direction we use the notation $\xx = (\rrho, x_n)$ with $\rho = |\rrho| = \big(\sum_{j=1}^{n-1} x_j^2 \big)^{1/2}$. 
$L$ denotes the elliptic operator $L := a \D + \bb \cdot \na + c$ with coefficients $a$, $\bb$, $c$ satisfying the regularity requirements 
$a$, $\na a$, $\rho \bb$, $\rho c \in \bigcap_{i=1}^k C^0(\Go_i \ti [0,T]) \cap C^0 (\Gh \ti [0,T])$ and $a \in C^0 (\Go \ti [0,T])$ with $a \geq a_0 >0$. So, $\bb$ and $c$ are possibly unbounded at $S$ and need not be continuous over $\Ga$, whereas $a$ is supposed to be continuous in all $G$. 
Functions $u =u(\rrho, z, t)$ then satisfy the system of inequalities
\be \label{5.2}
\le.
\ba{cl}
\ds \pa_t u \leq L u  & \mbox{ in } G_i \ti (0,T) \;\; \mbox { for } i=1, \ldots,k , \\[.5em]
\quad 0 \leq L u  & \mbox{ in } \Gh \ti (0,T) , \\[.5em]
u \mbox{ and } \na u \mbox{ continuous } & \mbox{ in } \R^n \ti (0,T) , \\[.5em]
u(\rrho,\cdot, \cdot) \leq 0 & \mbox{ for } \rho \ra 0 , \\[.5em]
u(\rrho, z, \cdot) \leq 0 & \mbox{ for } |\xx| \ra \infty, \\[.5em]
u(\cdot ,\cdot ,0) = u_0 & \mbox{ on } G \ti \{t=0\}.
\ea
\ri\}
\ee
in a classical sense if all derivatives exist as continuous functions, i.e.\ $u \in \bigcap_{i=1}^k C_1^2(G_i \ti (0,T)) \cap C^2(\Gh \ti (0,T))$. For such functions we have the following result
\begin{Theorem}[Maximum principle]
For functions $u \in (\Go \ti [0,T])$ with $T>0$, which satisfy (\ref{5.2}) with $c \leq 0$ in a classical sense, holds 
$$
\sup_{\R^n\ti(0,T]} u\, \leq\, \max_{\Go} \, u_0^+ .
$$
\end{Theorem}
Here, $u_0^+$ means the positive part of $u_0$, i.e.\ $u_0^+ (\xx) := \max \{u_0 (\xx), 0\}$, and $u \in C(\Go \ti [0,T])$ implies that $u$ takes continuously its initial value $u_0$ on $G$. A proof of theorem 3 is given in Appendix B; we make here only a few comments: our result differs from similar ones in (Backus 1957) or (Ivers \& James 1984) in that several separate conducting regions are admitted. This allows us to consider discontinuous flows, which typically have no classical solutions in all $G$. Theorem 3 states, in case of a positive initial maximum, that $u$ can never exceed this value and, in case of a negative initial maximum, that $u$ can never exceed zero.\footnote{We take the opportunity to correct the maximum principle in (Kaiser 2007), where we missed the case of a negative initial maximum: in lemma 1 $q_0$ should be replaced by $q_0^+$. All consequences we have drawn there from lemma 1 remain untouched by this correction.} Theorem 3 holds likewise in the bounded case, where (\ref{5.2})$_{2,5}$ are replaced by the boundary condition $u|_{\pa G \ti (0,T)} \leq 0$.

In this section it is convenient to use polar coordinates $(r,\th)$ in the meridional plane in the form $\rho =r \sin \th$, $z = r\cos \th$ with $r>0$ and $0<\th <\pi$. The symmetry axis is then represented by $\th = 0$ and $\pi$, $r> 0$. Equation (\ref{2.5}) for the meridional scalar $M = M(r,\th,t)$ takes then the form
\be \label{5.3}
\pa_t M - \eta \Big(\pa_r^2 M + \frac{\sin \th}{r^2}\, \pa_\th \Big(\frac{1}{\sin \th}\, \pa_\th M \Big)\Big) + v_r\, \pa_r M + \frac{1}{r}\, v_\th\, \pa_\th M = 0
\ee
and the axis condition (\ref{2.19}) reads
\be \label{5.4}
M(\cdot, \th, \cdot) = O(\sin^2 \th) \qquad \mbox{ for } \th \ra 0 \mbox{ or } \pi .
\ee
Obviously, $(r, \th, \phi)$ constitute spherical coordinates in $\R^3$.

Let us now consider the piecewise constant (discontinuous) velocity field $\vv$ in $B_1 \subset \R^3$:
\be \label{5.5}
\left.
\ba{c}
\quad\;\, v_r = \left\{
  \begin{array}{ll}
   \, - 3\,  c \qquad \mbox{ for } 1/3 < r < 2/3  \\
   \quad  3\,  c \qquad \mbox{ for } 2/3 < r < 1    
  \end{array} \right. , \quad 0 < \th < \pi , \\[1.5em]
\ds \frac{v_\th}{ r \sin \th} = \left\{
  \begin{array}{ll}
   \, - 9\,  c \qquad \mbox{ for } 0  < \th < \pi/2 \\
   \quad  9\,  c \qquad \mbox{ for } \pi/2 < \th < \pi 
  \end{array} \right. , \quad 1/3 < r < 1 , 
\ea \quad t \geq 0 \quad \right\}
\ee
with some constant $c>0$; otherwise $\vv$ is zero, in particular, $\vv = 0$ in $B_{1/3}$ and $v_\phi \equiv 0$.  Let $M$ be a solution of the meridional problem (\ref{2.14}) with this flow field, with $\eta =1$ and with initial value $M_0 \geq 0$. In general, $M$ will not be a classical solution in $G = B_1$, but $M$ as well as $-M$ will satisfy the premises of theorem 3, especially (\ref{5.2}) with $B_1$ subdivided into $B_{1/3}$ and four spherical half-shells. Applying theorem 3 on $-M$ then yields 
\be \label{5.6}
\inf_{\R^3\ti(0,T]} M\, \geq\, \min_{\xx \in \Bo_1} \{ \min \{M_0 (\xx), 0\}\} = 0
\ee
for any $T> 0$.

In the following we construct a {\em subsolution} $\Mu$ of (\ref{5.3}) (i.e.\ the left-hand side in (\ref{5.3}) is non-positive) with decay rate $d_m$ on the half-annulus $1/3 < r <1$, $0 <\th <\pi$ satisfying (\ref{5.4}) and the boundary condition $\Mu|_{r = 1/3} = \Mu|_{r=1} = 0$. In $\R^3$ this region corresponds to the spherical shell $S\!S:= B_1 \setminus \Bo_{1/3}$. The function $\Mu - M$ then satisfies on $S\!S$ the bounded version of (\ref{5.2}), in particular, we have by (\ref{5.6}), $(\Mu - M)|_{\pa S\!S} \leq 0$. So, choosing $\Mu_0 \leq M_0$, theorem 3 yields $\sup_{S\!S \ti (0,T]} (\Mu - M) \leq 0$ for any $T>0$ or, equivalently, $M(\cdot, t) \geq \Mu(\cdot, t)$ for any $t\geq 0$, i.e.\ $d_m$ is an upper bound on the decay rate of $M$.

To obtain a suitable subsolution $\Mu$ we make the ansatz
\be \label{5.7}
\Mu (r,\th, t) := f(r)\, g(\th)\, e^{- d_m t}
\ee
with non-negative functions $f$ and $g$, and a constant $d_m >0$. The construction of $f$ and $g$ is based on the auxiliary function 
\be \label{5.7a}
h(s) := e^{- c\, s/2} \sinh \sqrt{(c/2)^2 - d}\, s , \qquad 0<s<1 ,
\ee 
where the parameters $c$ and $d$ are related by
$$
\sqrt{(c/2)^2 - d}\, \coth \sqrt{(c/2)^2 - d} = c/2 ,
$$
which defines a function $d = D(c)$ that behaves asymptotically like 
\be \label{5.8} 
d = D(c) \sim c^2 e^{- c} \quad \mbox{ for } c \ra \infty .
\ee
$h$ is the unique solution of the boundary value problem 
\be \label{5.9}
\frac{\rd^2}{\rd s^2}\, h + c\, \frac{\rd}{\rd s}\, h + d\, h = 0 , \qquad h(0) = 0 , \quad \frac{\rd}{\rd s}\, h (1)= 0 .
\ee
Thus, with the substitution $t := 1 - s$, $h$ satisfies the inequality 
$$
(1- t^2) \frac{\rd^2}{\rd t^2}\, h - c (1-t^2) \frac{\rd}{\rd t}\, h + d\, h \geq 0 , \qquad 0< t< 1 ,
$$
or, after the further substitution $t := \cos \th$, 
$$
\sin \th\, \frac{\rd}{\rd \th} \Big(\frac{1}{\sin \th}\, \frac{\rd}{\rd \th}\, g_1 \Big) + c \sin \th\, \frac{\rd}{\rd \th}\, g_1 + d\, g_1
\geq 0 , \qquad 0 < \th < \frac{\pi}{2} ,
$$
where $g_1 (\th) := h(t)$. At the boundaries we have $g_1(0) = \rd/\!\rd \th\, g_1 (\pi/2) = 0$. Introducing $v_\th$ from (\ref{5.5}) and observing that $\rd/\!\rd \th\, g_1 \geq 0$ we have, finally,
\be \label{5.10}
\sin \th\, \frac{\rd}{\rd \th} \Big(\frac{1}{\sin \th}\, \frac{\rd}{\rd \th}\, g_1 \Big) -r v_\th\, \frac{\rd}{\rd \th}\, g_1 + d\, g_1
\geq 0 , \qquad 0 < \th < \frac{\pi}{2} , \quad \frac{1}{3} < r <1 .
\ee
By reflection at $\th = \pi/2$ we find that $g_2 (\th) := g_1 (\pi - \th)$ satisfies inequality (\ref{5.10}) in the region $\pi/2 < \th < \pi$, $1/3 < r < 1$.

Concerning the radial part we start from (\ref{5.9}) with $d$ replaced by $\dti := d_m /9 - d$ and find by the substitution 
$r := (1+s)/3$ for $f_1 (r) := h(s)$ the inequality
\be \label{5.11}
r^2 \frac{\rd^2}{\rd r^2}\, f_1 - r^2 v_r\, \frac{\rd}{\rd r}\, f_1 + r^2 d_m f_1 - d\, f_1 \geq 0 , \qquad 1/3< r < 2/3 
\ee
and the boundary conditions $f_1 (1/3) = \rd/\!\rd r f_1 (2/3) = 0$. An analogous calculation yields $f_2 (r)$ satisfying (\ref{5.11}) on $2/3 < r <1$ with boundary conditions  $\rd/\!\rd r f_2 (2/3) = f_2 (1) = 0$.

The functions $f_{1,2}$ and $g_{1,2}$ constitute now the subsolution (\ref{5.7}) on their respective domains: (\ref{5.10}) and (\ref{5.11})
imply (\ref{5.3}) (as inequality) with $\eta \equiv 1$ and one easily checks $C^1$-smoothness over the subdomains, the asymptotic condition (\ref{5.4}) and the boundary conditions $\Mu(1/3, \cdot, \cdot) =  \Mu(2/3, \cdot, \cdot) = 0$. For large flow amplitudes $c$, the quantities $d$ and $\dti$ and hence $d_m$ obey (\ref{5.8}), which manifests the (exponentially) slow decay of $M$.

To demonstrate that slow decay is not a peculiarity of discontinuous flow fields let us consider one more velocity field, viz.
\be \label{5.A1}
\le.
\ba{l}
\ds v_r = 3 c\; \frac{1}{2} \big(1 + \tanh (\de (r -1/3))\big) \tanh (\de (r - 2/3)) \tanh (\de (r- 1/10)) \tanh (\de (1-r)) , \\[0.5em]
v_{\th} = 9 c \; r \sin \th\, \tanh \big((\de/4) (\th - \pi/2)\big) ,
\ea
\ri\}
\ee
which is a smoothed version of (\ref{5.5}). The dynamo equation has been solved
numerically with this flow field and $\eta =1$ in the spherical shell $\SSS := \{ \xx : 1/10 < |\xx|
<1\}$. Both regions $|\xx|<1/10$ and $|\xx|>1$ are assumed to be vacuum. The inner vacuum
sphere simplifies the numerical procedure because it excludes the
coordinate singularity from the computational volume. The factors $\de$ and
$\de/4$ in (\ref{5.A1}) are chosen this way in order to smooth the discontinuity
in the original flow field (\ref{5.5}) over roughly the same length in radial and
latitudinal directions. 

\begin{figure}[h]
\begin{center}
\includegraphics[width=11cm]{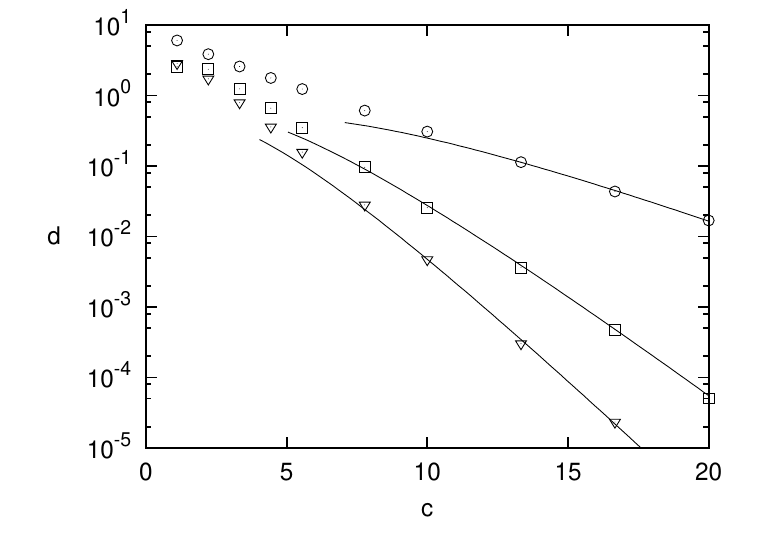}
\end{center}
\caption{Decay rate $d$ of the magnetic field as a function of the flow amplitude $c$ for $\de=4$
(circles), 8 (squares) and 20 (triangles). The solid lines are the following exponential
fits to the data: $d \sim c^2 e^{-0.41c}$ (circles), $d \sim c^2 e^{-0.76 c}$ (squares) and $d \sim c^2 e^{-0.96 c}$ (triangles).
}
\label{fig1}
\end{figure}

The numerical code implements a standard time
integration procedure with a spatial discretization in which the magnetic field
is represented in terms of poloidal and toroidal scalars;
both scalars are decomposed into Chebychev polynomials for the
radial direction and spherical harmonics for the angular variables. Because we
are only interested in axisymmetric solutions, the code is used with a
truncation level such that only the axisymmetric spherical harmonics, i.e.\
Legendre polynomials, are retained. The time stepping employs second order
Adams-Bashforth and Crank-Nicolson schemes for the induction and diffusion
terms, respectively. The magnetic energy $E_B=\frac{1}{2} \int_{\SSS} \BB^2 \rd x$
is recorded during the time integration.
The calculation is started from an arbitrary superposition of dipolar and
quadrupolar fields and is continued until, after initial transients, the time
dependence of $E_B$ is close to an exponential decay of the form
$E_B (t) \sim e^{-2dt}$, from which the decay rate $d$ of the magnetic field is
deduced. Figure \ref{fig1} shows $d$ as a function of $c$ for $\de$ between 4
and 20 and a resolution of 128 Chebychev polynomials and Legendre polynomials of
degree up to 128. The solid lines in that figure demonstrate that the functional dependence of $d$ on $c$ 
is well described by $d \sim c^2 e^{-const\ti c}$ for $c$ large enough. Moreover, for 
moderate smoothing (triangles) the exponential dependence is compatible with that 
predicted by (\ref{5.8}). 

Note, finally, that the flows (\ref{5.5}) or (\ref{5.A1}) are special in that they exhibit sources (at the surface of discontinuity in (\ref{5.5})) and sinks (at the boundaries of the flow region) of the fluid. In incompressible and weakly compressible fluids slow decay cannot occur as demonstrated in section 6. Likewise, simpler flows, i.e.\ one-dimensional flows as considered in appendix C, do not allow slow decay even when discontinous or violating mass conservation. So, the precise fluid or flow property responsible for slow decay is an open question.
\section{Incompressible and weakly compressible fluids}
When the flow field $\vv$ is such that $\na \cdot \vv = 0$ and the diffusivity $\eta$ is constant, much larger decay rates can be obtained than without these restrictions. In this case the flow field drops from the energy balance and meridional and azimuthal lower decay rate bounds are obtained as extremal values of certain variational problems for the meridional and azimuthal fields, respectively. These bounds  depend neither on the form nor on the amplitude of the flow and are comparable in size to those of free decay. To extend the method to the weakly compressible case we make use of the equation of mass conservation 
\be \label{6.1}
\pa_t \rh + \na \cdot (\rh\, \vv) = 0 ,
\ee
which relates the violation of the divergence-constraint $\na \cdot \vv = 0$ to the variation of the mass density $\rh (\xx, t)$. $\rh$ is assumed to be a smooth function, at least $\rh \in C^2 (G \ti \R_+)$, with positive bounds $\ru \leq \rh (\xx, t) \leq \ro$.

We start in the meridional case with an ansatz that follows Backus (1957) and that leads to additional terms in the energy balance. Using, however, a different dynamic variable than in (Backus 1957) our additional terms depend only on the (relative) variation of density and diffusivity. Controlling these terms by a smallness assumption the variational problem still yields decay rate bounds from below comparable to those for constant density and diffusivity. 

Multiplying (\ref{2.14})$_1$ by $\rh M$ and (\ref{6.1}) by $M^2 /2$, adding and integrating over the regularized domain $G_\ep$ introduced in section 4, integrating by parts and using the boundary condition $\nn \cdot \vv = 0$ on $S_\ep$ yields
\be \label{6.2}
\ba{rl}
\ds \frac{1}{2}\,\frac{\rd}{\rd t} \int_{G_\ep} \rh M^2 \rd x \!\!\!
&\ds = \int_{G_\ep} \eta \rh \Big(M \D M - \na \cdot \Big( \frac{\na \rho}{\rho} M^2 \Big) \Big) \rd x
- \frac{1}{2}\int_{G_\ep} \na \cdot (\rh M^2 \vv )  \rd x \\[1em]
& \ds = - \int_{G_\ep} |\na (\ka^{1/2} M)|^2 \rd x + \int_{G_\ep} \Big( |\na \ka^{1/2} |^2 + \frac{\na \rho}{\rho} \cdot \na \ka \Big) M^2 \rd x \\[1em]
& \quad \ds + \int_{S_\ep \cup C_\ep} \ka M \Big( \nn\cdot \na M - \frac{\na \rho}{\rho}\cdot  \nn\, M \Big) \rd s + \frac{1}{2} \int_{C_\ep} \rh\, \na\rho \cdot \vv\, M^2 \rd s .
\ea
\ee
We introduced here the abbreviation $\ka := \eta \rh$, whose local variation appears in (\ref{6.2}) in the form\footnote{Note that axisymmetry of the smooth function $\ka$ ensures a finite variation $\Vk$ also on the symmetry axis.}
$$
\Vk := \frac{|\na \ka^{1/2} |^2}{\ka} + \frac{\na \rho \cdot \na \ka}{\rho \ka} = 
\Big|\frac{\na (\eta \rh)}{2 \eta \rh}\Big|^2  + \frac{1}{\rho}\, \frac{\pa_\rho (\eta \rh)}{\eta \rh} \ .
$$
To cope with the boundary terms in (\ref{6.2}) it is convenient to introduce an extension $\kti$ of $\ka$ onto $\R^3$ with the properties: $\kti \in C^0 (\R^3\ti \R_+) \cap C^2 (G \ti \R_+) \cap C^2 (\Gh \ti \R_+)$, $\sup_{\R^3} \Vkti = \max_{\Go} \Vk =: \Vk_{max}$ and $\na \ka = 0$ outside some ball $B_{\Rti}$. In the case $\ka|_{\pa G} = const$ we take simply $\Rti = R$, where $R$ is the radius of the smallest ball enclosing $G$, otherwise $\Rti = (1 + \pi) R$ is sufficient to construct an extension of the above kind.
Using this extension an analogous calculation yields in $\Gh_\ep := \{\xx \in \Gh : \rho(\xx) > \ep\}$: 
\be \label{6.4}
\ba{c}
\ds 0 = \int_{\Gh_\ep} \kti \Big(M \D M - \na\cdot \Big( \frac{\na \rho}{\rho}\, M^2 \Big)\Big) \rd x 
= - \int_{\Gh_\ep} |\na(\kti^{1/2} M)|^2 \rd x + \int_{\Gh_\ep} \Vkti\, \kti\, M^2 \rd x \\[1em]
\ds -  \int_{S_\ep} \kti M\Big( \nn\cdot \na M - \frac{\na \rho}{\rho}\cdot \nn\, M \Big) \rd s - \int_{\Ch_\ep} \kti M \Big(\na \rho \cdot \na M - \frac{1}{\rho}\, M \Big) \rd s  ,
\ea
\ee
where we used the asymptotic conditions (\ref{2.14})$_5$ and (\ref{2.18}) and the notation $\Ch_\ep := \{\xx \in \Gh : \rho(\xx) = \ep\}$. Summing up (\ref{6.2}) and (\ref{6.4}) and letting $\ep \ra 0$ while observing (\ref{2.14})$_{3,4}$ yields, finally, the energy balance
\be \label{6.5}
\frac{1}{2}\, \frac{\rd}{\rd t} \int_{G} \rh\, M^2 \rd x = -  \int_{\R^3} |\na (\kti^{1/2} M)|^2 \rd x + \int_{B_{\Rti}} \Vkti\, \kti\, M^2 \rd x .
\ee
Equation (\ref{6.5}) suggests to consider the variational problem
\be \label{6.6}
\inf_{0 \neq f \in S_m} \frac{\int_{\R^3} |\na f|^2 \rd x}{\int_G f^2 \rd x }\; =: \la_1 ,
\ee
where $S_m$ means the set of axisymmetric functions that satisfy (\ref{2.14})$_{3,4,5}$. 
The explicit determination of $\la_1 = \la_1 (G)$ is for general $G$ no easy task, not even in balls since minimizing functions satisfying the axis-condition (\ref{2.14})$_4$ cannot expected to separate in spherical nor in cylindrical coordinates. Bounds, however, can easily be obtained: lower bounds by enlarging the variational set $S_m$ (e.g.\ by removing conditions on its elements) and upper bounds by inserting any test functions. Moreover, as is obvious from (\ref{6.6}), $\la_1$ is a monotonous function of $G$, i.e., in particular, $B_r \subset G \subset B_R$ implies $\la_1 (B_R) \leq \la_1 (G) \leq \la_1 (B_r)$.

A lower bound $\lau$ is obtained by removing the conditions of axisymmetry and (\ref{2.14})$_4$ and by replacing $G$ by $B_R$. In this case (\ref{6.6}) becomes equivalent to the well-known poloidal variational problem (without zero-mean condition), where the function 
$$
f_1 (\xx) := 
\le\{
\ba{ll}
j_0 (\pi |\xx| / 2 R ) & \text{ in } B_R , \\[.1em]
j_0 (\pi / 2 )\, R/|\xx| & \text{ in } \Bh_R  
\ea
\ri.
$$
minimizes (\ref{6.6}) with the result $\lau = (\pi/ 2 R)^2$ (cf.\ Backus 1958). Note that $f_1$ is in fact axisymmetric but does not satisfy the axis-condition (\ref{2.14})$_4$. An uppper bound $\lao$ is obtained by the following test function that respects the axis-condition: 
$$
f_2 (\xx) := 
\le\{
\ba{ll}
j_0 (p |\xx| / r )\, (|\xx|/r)^2 \sin^2 \theta & \text{ in } B_r , \\[.1em]
j_0 (p )\, (r/|\xx|) \sin^2 \theta & \text{ in } \Bh_r .
\ea
\ri.
$$
Here, $j_0$ denotes the zeroth spherical Bessel function, $\theta$ is the angle between $\xx$ and the symmetry axis and $p$ is a free parameter. Inserting $f_2$ into (\ref{6.6}) we find a minimum value $\lao \approx 8.85/r^2$ for $p \approx 2.6$.\footnote{Note that $f_2$ is an admissible test function although it is not differentiable at $S_r$ for this parameter value. 
} In summary we have the bounds
\be \label{6.7}
\pi^2/4 R^2 \leq \lambda_1 (G) \lesssim 8.85/r^2 ,
\ee
where $0<r<R$ are such that $B_r \subset G \subset B_R$.

Now, estimating the right-hand side in (\ref{6.5}) by (\ref{6.6}) we obtain 
$$
\frac{\rd}{\rd t} \int_{G} \rh\, M^2 \rd x = -2 \big(\lambda_1 (B_{\Rti}) - \Vk_{max}\big)  \int_{B_{\Rti}} \kti\, M^2 \rd x \leq -2 d_m^{wc}  \int_G \rh \, M^2 \rd x , 
$$
which implies exponential decay with rate $d_m^{wc} := \eta_0 (\la_1 (B_{\Rti} ) - \Vk_{max})$ provided $\Vk_{max} < \la_1 (B_{\Rti} )$. 

Our results imply the incompressible case with constant diffusivity, in which case we have $d_m^{wc} := \eta \la_1 (G)$. Especially for balls our bounds (\ref{6.7}) can be compared to those given previously by other authors: (\ref{6.7}) corroborates Backus' lower bound but dismisses Braginskii's claim that $\la_1 (B_1) = \pi^2$. 


In the azimuthal case the following ansatz turns out to be successful: multiply (\ref{2.15})$_1$ by $\cA/ \rh$ and subtract (\ref{6.1}) multiplied by $\cA^2 /2 \rh^2$. Integrating over $G_\ep$, integrating by parts and using (\ref{2.15})$_2$ yields by a similar calculation as in section 4:
$$
\ba{rl}
\ds \frac{1}{2}\,\frac{\rd}{\rd t} \int_{G_\ep} \frac{\cA^2}{\rh}\, \rd x \!\!\!
& \ds = - \int_{G_\ep} \eta \, |\na (\rh^{-1/2} \cA)|^2\, \rd x + \int_{G_\ep} \Big(\eta \big|\na \rh^{-1/2} \big|^2 + \frac{1}{\rh^2}\,\frac{\na \rho}{\rho} \cdot \na (\eta \rh) \Big) \cA^2 \rd x \\[1em]
& \quad \ds - \int_{C_\ep} \frac{\eta}{\rh}\, \cA \Big(\na \rho\cdot \na \cA - \frac{1}{\rho}\, \cA \Big) \rd s + \frac{1}{2} \int_{C_\ep} \frac{1}{\rh}\, \na\rho \cdot \vv\, \cA^2 \rd s .
\ea
$$
For simplicity we have left out the meridional source term. With the abbreviation
$$
\Ver := \frac{\eta}{\eta_0}\, \rh\, \big| \na \rh^{-1/2}\big|^2 + \frac{\na \rho \cdot \na (\eta \rh)}{\rho \eta_0 \rh} = 
\frac{\eta}{\eta_0} \bigg(\Big|\frac{\na \rh}{2 \rh}\Big|^2  + \frac{1}{\rho}\, \frac{\pa_\rho (\eta \rh)}{\eta \rh} \bigg) 
$$
we then obtain in the limit $\ep \ra 0$ the azimuthal energy balance
\be \label{6.8}
\frac{1}{2}\, \frac{\rd}{\rd t} \int_{G} \rh^{-1} \cA^2 \rd x = -  \int_{G} \eta \, |\na (\rh^{-1/2} \cA)|^2 \, \rd x 
-2 \pi \int_L \eta\, \rh^{-1} \cA^2 \rd z + \eta_0 \int_{G} \Ver\, \rh^{-1} \cA^2 \rd x ,
\ee
which suggests to consider the following variational problem:
\be \label{6.9}
\inf_{0 \neq g \in S_a} \frac{\int_{G} |\na g|^2 \rd x + 2\pi \int_L g^2 \rd z}{\int_G g^2 \rd x }\; =: \mu_1 ,
\ee
where $S_a$ means the set of axisymmetric differentiable functions vanishing on $\pa G$. A lower bound $\muu$ on (\ref{6.9}) is obtained by neglecting the line integral. In this case the function $g_1 (\xx) := j_0 (\pi |\xx|/r)$ is well-known to minimize (\ref{6.9}) in the ball $B_R$ with the result $\muu = (\pi/R)^2$ (cf.\ Backus 1958). Computing the full expression (\ref{6.9}) with $g_1$ yields the upper bound $\muo = (\pi^2 + 2 \pi Si(2 \pi))/R^2$, where $Si$ denotes the sine integral. In summary we have the bounds 
\be \label{6.10}
\pi^2/R^2 \leq \mu_1 (G) \lesssim 18.78/r^2 ,
\ee
where $0<r<R$ are such that $B_r \subset G \subset B_R$. As in the meridional case corroborates (\ref{6.10}) Backus' lower bound (derived by a completely different method) and dismisses Braginskii's claim that $\mu_1 (B_1) \approx 20.19$.  

By (\ref{6.9}) and with $\Ver_{max} := \max_{\Go} \Ver$ we can estimate the right-hand side in (\ref{6.8}) to obtain 
$$
\frac{\rd}{\rd t} \int_{G} \rh^{-1} \cA^2 \rd x = -2 \eta_0 (\mu_1 (G) - \Ver_{max}) \int_G \rh^{-1} \cA^2 \rd x , 
$$
which implies exponential decay with rate $d_a^{wc} := \eta_0 (\mu_1 (G) - \Ver_{max})$ provided $\Ver_{max} < \mu_1 (G)$. 

Pointwise decay can again be obtained by means of theorem 1. In the azimuthal case without meridional field pointwise decay holds with almost the weakly compressible decay rate $d_a^{wc}$ since the five-dimensional energy (which is required by theorem 1) is dominated by the three-dimensional energy. This is not so in the meridional case. Only when using additional information, five-dimensional energy decay (with a lowered decay rate) can be established: splitting the conducting region $G$ into $G_\ep$ and its complement and using (\ref{3.1}) we can estimate  
$$
\ba{rl}
\ds \int_{G_5} \cM(\xx, t)^2 \rd x \!\!\! &\ds = \frac{|S_3|}{2 \pi} \int_{G} \Big(\frac{1}{\rho^2} M(\xx,t)\Big)^2 \rho^2 \rd x \\[.8em]
&\ds \leq \frac{|S_3|}{2 \pi}\, \frac{1}{\ep^2} \int_{G_\ep } M(\xx,t)^2 \rd x + \int_{G \setminus G_\ep} g(\rho)^2 \frac{1}{\rho^2}\, \rd x 
\leq \frac{\Cti}{2 \ep^2}\, e^{-2 d_m^{wc} t} + \frac{\Cti}{2}\, \ep^2 .
\ea
$$
Setting $\ep := \exp (- d_m^{wc}\, t/2)$ we thus have 
$
\int_{G_5} \cM(\xx, t)^2 \rd x \leq \Cti \, e^{- d_m^{wc} t} 
$
and hence by (\ref{3.12}) pointwise decay of the meridional field with half the weakly compressible decay rate (which is supposedly not optimal). 

Finally, using the information of the previous paragraph, the azimuthal energy balance {\em with} the meridional sorce term can also be handled. A calculation analogous to that in section 4 just replaces the decay rate $d_a^{wc}$ by $\min \{d_a^{wc} - \ep\, ,\, d_m^{wc}/2 \}$.

%
%
\setcounter{section}{0}
\renewcommand{\thesection}{\Alph{section}}
\app{A}
This appendix expounds the necessary changes in the proof of (Lortz et al.\ 1984, theorem 2) to be applicable to theorem 2. First we rewrite system (\ref{4.1}) in $\R^5$. In view of (\ref{2.17}) one obtains
\be \label{A.1}
\le.
\ba{cl}
\ds \pa_t P - \na \cdot ( \eta \na P) +\na \cdot (\vv_m^{(5)} P) + c\, P = 0 & \mbox{ in } G_5 \ti (0,T) , \\[.5em]
\nn \cdot \na P = 0 & \mbox{ on } \pa G_5 \ti (0,T) , \\[.5em]
P(\cdot\, ,0) = P_0  > 0 & \mbox{ on } G_5 \ti \{t=0\}
\ea
\ri\}
\ee
with $c:= -2 (\pa_\rho \eta /\rho + v_\rho /\rho)$ and arbitrary $T>0$. System (\ref{A.1}) differs from system (\ref{2.2}) in (Lortz et al.\ 1984)\footnote{Equation numbers referring to (Lortz et al.\ 1984) are henceforth marked by the suffix `LMS', thus, `(LMS2.2)'.} by the zeroth-order term $c\, P$. Classical solutions are guaranteed for both systems by the sufficient conditions $\pa G \in C^3$, $\eta$ and $\vv_m^{(5)} \in C_1^2 (\Go_5 \ti [0,T])$ with $\nn \cdot \vv_m^{(5)} |_{\pa G \ti [0, T]} = 0$, $c \in C_1^1(\Go_5 \ti [0,T])$ and $P_0 \in C^3 (\Go_5)$ with $\nn \cdot \na P_0 |_{\pa G} = 0$, which imply the precise H\"older conditions formulated, e.g., in (Ladyzenskaja et al.\ 1968, chap.\ IV, theorem 5.3). The proof of (Lortz et al.\ 1984, theorem 2) refers to such classical solutions.

The governing equation (LMS2.2a) enters the proof two times, viz.\ in the derivations of the inequalities (LMS3.1) and (LMS3.12). We show in the following that the zeroth-order term in (\ref{A.1})$_1$ does not invalidate these inequalities, it merely modifies the bounds that depend then, additionally, on $c$.

Modified proof of (LMS3.1): Multiplying  (\ref{A.1})$_1$ by $P^{\ga -1}$ with $\ga \ne 0,1$, integrating over $G$, integrating by parts and using the boundary conditions for $P$ and $\vv_m^{(5)}$ yields  
\be \label{A.2}
\ba{rl}
\ds \frac{1}{\ga}\,\frac{\rd}{\rd t} \int_{G} P^\ga \, \rd x\!\!\! &= \ds \int_G \pa_t P \, P^\ga\,\rd x = \int_{G} \big\{ \na \cdot (\eta \na P - \vv_m^{(5)} P ) P^{\ga -1} - c\, P^\ga \big\} \rd x\\[1em]
& = \ds \int_{G} \big\{ (1-\ga) (\eta P^{\ga -2} |\na P|^2  - P^{\ga -1} \vv_m^{(5)} \cdot \na P ) - c\, P^\ga \big\} \rd x .
\ea
\ee
With the bounds $\eta \geq \eta_0$, $|\vv_m^{(5)}| \leq K$ and $|c|\leq K$, and by Young's inequality, (\ref{A.2}) can be estimated as follows:
$$
\ba{rl}
\ds \frac{1}{\ga (\ga -1)}\,\frac{\rd}{\rd t} \int_{G} P^\ga \, \rd x\!\!\! &\leq \ds - \eta_0 \int_G P^{\ga -2} |\na P|^2 \rd x  + K \int_G P^{\ga -1} |\na P|\, \rd x + \frac{K}{|\ga -1|} \int_G P^\ga \rd x \\[1em]
&\leq \ds - \frac{3}{4}\, \eta_0 \int_G P^{\ga -2} |\na P|^2 \rd x  + \Big(\frac{K^2}{\eta_0} + \frac{K}{|\ga -1|} \Big) \int_G P^\ga \rd x .
\ea
$$
Finally, multiplying by $\phi (t)$, integrating over $[t_1, t_2]$ and integrating by parts yields (LMS3.1) with the constant $K^2/\eta_0$ replaced by $K^2/\eta_0 + K/|\ga -1|$. In the course of proof the parameter $\ga$ takes infinitely many values; however, always holds $|\ga -1|^{-1} \leq n+1$ if $n \geq 3$, in particular, $|\ga -1|^{-1} \leq 6$ in $\R^5$ (Lortz et al.\ 1984, p.\ 688 and p.\ 690). Thus, (LMS3.1) holds with an enlarged constant.

Modified proof of (LMS3.12): Multiplying  (\ref{A.1})$_1$ by $P^{-1}$ and integrating over $G$ yields analogously: 
$$
\ba{rl}
\ds \frac{\rd}{\rd t} \int_{G} \log P\, \rd x\!\!\! &= \ds \int_{G} \big\{ \na \cdot (\eta \na P - \vv_m^{(5)} P ) P^{-1} - c \big\} \rd x\\[1em]
& = \ds \int_{G} \eta |\na \log P|^2 \rd x - \int_G \na (\log P) \cdot \vv_m^{(5)} \, \rd x  - \int_G c\, \rd x \\[1em]
& \geq \ds \frac{1}{2}\, \eta_0 \int_G |\na \log P|^2 \rd x - \frac{1}{2}\, \frac{K^2}{\eta_0} \, |G| - K |G| ,
\ea
$$ 
which is (LMS3.12) with the constant $K^2 |G|/(2 \eta_0)$ enlarged by $K|G|$.
\app{B}
Proof of theorem 3:  Let $M$ denote $\sup_{\R^n\ti(0,T]} u$, which is finite because of (\ref{5.2})$_5$. The case $M \leq 0$ is trivial: by continuity we have $\sup_{\Go} u_0 \leq 0$ and hence $M\leq 0 = \max_{\Go} u_0^+$.

To prove the case $M>0$ we make use of the elliptic maximum principle and the elliptic boundary derivative theorem for classical solutions
as formulated in (Protter \& Weinberger 1984, theorems 6-8 at p.\ 64ff) and the parabolic maximum principle for weak solutions as formulated in (Lieberman 1996, theorem 6.25 and corollary 6.26 at p.\ 128). We show first that $\sup_{\Gh} u(\cdot, t) < \max_{\Go} u(\cdot, t)$ for any $t \in (0,T]$ and, second, that $\max_{\Go \ti [0,T]} u$ is taken at $t=0$, which proves our assertion. 

\noi 1) Let $t_0 \in (0,T]$ and $\Mh_{t_0} := \sup_{\Gh} u(\cdot, t_0)$. Because of (\ref{5.2})$_{4,5}$ we find a ball $B$ with complement $\Bh$ and a solid cylinder $\SC$ around the symmetry axis with complement $\SCh$ such that $u(\cdot, t_0) < \Mh_{t_0}$ in $\Bh \cup \SC$. Applying the elliptic maximum principle on the bounded and regularized domain $\Gh \cap B \cap \SCh$ we find $\Mh_{t_0}$ be attained at some point $(\xx_0, t_0) \in \pa \Gh = \pa G$. Furthermore, as $u \neq const$ in $\Gh$ and $\pa G$ being smooth the boundary derivative theorem is applicable with the result
\be \label{B.1}
\nn \cdot \na u|_{(\xx_0, t_0)} < 0 ,
\ee
where $\nn$ denotes the exterior normal with respect to $G$ at $(\xx_0,t_0)$. By (\ref{5.2})$_3$ condition (\ref{B.1}) implies $u(\xx, t_0) > \Mh_{t_0}$ for some $\xx \in G$ and hence $M > \Mh_{t_0}$ for any $t_0$, i.e.\ 
\be \label{B.2}
M > u|_{\pa G  \ti (0,T]} \, .
\ee
\noi 2) As $u$ need not be a classical solution in all $G\ti (0,T)$ a parabolic maximum principle for {\em weak} solutions is better suited to complete te proof.  In order to check that $u$ satisfies (\ref{5.2})$_1$ in the weak sense of (Lieberman 1996, p.\ 100 above) we write $L$ in the form  
\be \label{B.3}
L = \na \cdot (a \na) + (\bb - \!\na a)\cdot \na + c
\ee
and choose again a solid cylinder $\SC$ with boundary $C$ such that $u < M$ in $\SC \cap G$, which implies, in particular, 
\be \label{B.4}
M > u|_{(C \cap G)  \ti (0,T]} \, .
\ee
Let $v \in C^1 ((\overline{\SCh \cap G}) \ti [0,T])$ be a non-negative testfunction that vanishes at $\pa (\SCh \cap G) \ti [0,T]$.  Integrating $v L u$ over $(\SCh \cap G_i) \ti (0,T)$, integrating the first term by parts and summing over $i = 1, \ldots, k$ reveals that $u$ satisfies indeed (\ref{5.2})$_1$ in $(\SCh \cap G) \ti (0,T)$ in the weak sense. Note that the regularity assumptions on $u$ and the coefficients are such that the boundary terms on $\Ga$ cancel each other; otherwise, boundedness of the coefficients in (\ref{B.3}) and the sign condition on $c$ are the only further prerequisites of the parabolic maximum principle. Therefore, $M$ is taken at the `parabolic boundary', which means by (\ref{B.2}) and (\ref{B.4}) that $M$ is taken at $t=0$.

The bounded version of theorem 3 is obviously proved by 2) and (\ref{B.2}), which holds now by virtue of the boundary condition $u|_{ \pa G \ti (0,T]} \leq 0$.
\app{C}
This appendix demonstrates that simpler flows than that considered in section 5, viz.\ purely radial or purely non-radial flows, even when discontinuous or violating mass conservation, do not exhibit slow decay, i.e.\ the meridional decay rates (even for large flow amplitudes)
do not drop significantly below the corresponding free decay rates. 
These results are obtained by constructing suitable supersolutions for the dynamo problem. 
A {\em supersolution} $\Mo$ (i.e.\ $-\Mo$ satisfies (\ref{5.2})) with decay rate $d_m$ that bounds a solution $M$ at $t=0$ does so for all time, hence $d_m$ provides a lower bound on the decay rate of $M$. This follows by theorem 3 applied on $M - \Mo$ and on $-M - \Mo$.

We consider first purely radial flows, i.e.\ $v_\th \equiv 0$, in an arbitrary conducting region $G$. Concerning $v_r$ the following bounds will play a role:
\be \label{5.12}
\left.
\ba{c}
- v_r \leq 2\, c_1\, \eta_0 /R \qquad \mbox{ for } 0 < r < R/2 , \\[.5em]
\quad v_r \leq 2\, c_2\, \eta_0 /R \qquad \mbox{ for } R/2 < r < R , 
\ea \quad 0 < \th < \pi , \quad t \geq 0 , \quad \right\}
\ee
where $\eta_0$ is a lower bound on $\eta$ and $R$ the radius of a ball enclosing $G$. For large flow amplitudes $c_1$, $c_2$ a supersolution is then given by  
\be \label{5.13}
\Mo_r (r,\th, t) := f(r)\, \sin^2 (\th)\, e^{- d_m^r t} ,
\ee
where $d_m^r := 8 \eta_0 /R^2$ and 
$$
f(x) := \left\{
\ba{ll}
\ds a_1 \sqrt{x}\, e^{-c_1 x} I_{3/2} (b_1 x) & 0 < x< 1/2 , \\[.3em]
\ds a_2 \sqrt{x}\, e^{c_2 x} (K_{3/2} (b_2 x) + a_3 I_{3/2} (b_2 x)) & 1/2 < x< 1 , \\[.3em]
\ds x^{-1} & x>1
\ea \right.
$$
with $x:= r/R$ and $b_i := (c_i^2 -d_i)^{1/2}$ $(i=1,2)$. $I_\nu$ and $K_\nu$ denote modified Bessel functions of order $\nu$ and $I'_\nu$ and $K'_\nu$ their derivatives. Thus $f(x)$ is a solution of 
\be \label{5.13a}
x^2 f'' - 2 (-1)^i c_i\, x^2 f' + d_i\, x^2 f - 2 f = 0 \qquad (i=1,2)
\ee
on $0 < x <1/2$ $(i=1)$ and $1/2 < x < 1$ $(i=2)$, respectively. To ensure $C^1$-smoothness we require $f'(1/2-) = 0$ and $f'(1/2+) = 0$, which yield
\be \label{5.14}
(1- c_1) I_{3/2} (b_1 /2) + b_1 I'_{3/2} (b_1 /2) = 0 
\ee
and 
\be \label{5.15}
(1+ c_2) (K_{3/2} (b_2 /2) + a_3 I_{3/2} (b_2 /2)) + b_2 (K'_{3/2} (b_2 /2) + a_3 I'_{3/2} (b_2 /2)) = 0 ,
\ee
respectively. $a_3$ is determined by the condition $f'(1-) = - f(1-)$ and $a_1$ and $a_2$ are chosen such that $f$ is continous. Relation (\ref{5.14}) determines a function $d_1 = D_1 (c_1)$ that for large arguments decreases monotonically to the asymptotic value $\lim_{c_1 \ra \infty} D_1 (c_1) = 8$.\footnote{This analysis has to be done with some care but it is elementary since $I_{3/2}$§ and $K_{3/2}$ can be expressed by elementary functions (cf.\ Abramowitz \& Stegun 1972, p.\ 443).} Similarly, (\ref{5.15}) determines a function $D_2$ with the same asymptotic property. Note that $f$ is monotonically increasing for $0 < x< 1/2$ and decreasing for $1/2 < x <1$, which allows us to use the bounds (\ref{5.12}). Thus, $\Mo_r$ satisfies (\ref{5.3}) (as inequality) in $B_R$ and hence in $G$. The remaining conditions in (\ref{5.2}) are easily checked. This verifies $\Mo_r$ to be a supersolution with decay rate $8 \eta_0 /R^2$, which differs from the free decay rate $\pi^2 \eta_0 /R^2$ by less than $20\%$. This bound cannot be improved since for a piecewise constant flow field in a ball according to (\ref{5.12})  with $c_1/c_2$ chosen such that $d_1 = d_2$, (\ref{5.13}) becomes an exact solution of the dynamo problem.

For purely non-radial flows holds  $v_r \equiv 0$. With the bounds on $v_\th$,
$$
\ba{c}
- r v_\th \leq c\, \sin \th\, \eta_0 \qquad \mbox{ for } 0 < \th < \pi/2 , \\[.2em]
\quad r v_\th \leq c\, \sin \th\, \eta_0 \qquad \mbox{ for } \pi/2 < \th < \pi , 
\ea \quad r < R , \quad t \geq 0 , 
$$
a supersolution reads now:
$$
\Mo_{nr} (r,\th, t) := f(r)\, g (\th)\, e^{- d_m^{nr} t} ,
$$
where $d_m^{nr} := \pi^2 \eta_0 /(4 R^2)$, $f(r):= \sin \pi r /(2 R)$ for $r< R$ and $f(r) := 1$ for $r >R$, and $g(\th) := h(1- \cos \th)$ for $0< \th < \pi/2$ and $g(\th) := h(1+ \cos \th)$ for $\pi/2 < \th < \pi$ with $h$ being the function (\ref{5.7a}). The verification of the conditions (\ref{5.2}) is left to the reader.
%
%
%
%
%
\subsection*{References}
\noi Abramowitz, M. \& Stegun, I. A., {\em Handbook of mathematical functions} (Dover Publications, New York 1972).

\vspace{1mm}\noi Backus, G., The axisymmetric self-excited fluid dynamo, {\em Astrophys.\ J.} {\bf 125}, 500--524 (1957).

\vspace{1mm}\noi Backus, G.E., A class of self-sustaining dissipative spherical dynamos, {\em Ann.\ Phys.} {\bf 4}, 372--447 (1958).

\vspace{1mm}\noi Backus, G.E. \& Chandrasekhar, S., On Cowling's theorem on the impossibility of self-maintained axisymmetric homogeneous dynamos, {\em Proc.\ Nat.\ Acad.\ Sci.\ USA} {\bf 42}, 105--109 (1956).

\vspace{1mm}\noi Braginskii, S.I., Self-excitation of a magnetic field during the motion of a highly conducting fluid, {\em Soviet Phys.\ JETP} {\bf 20}, 726--735 (1965). 


\vspace{1mm}\noi Cowling, T. O., The magnetic field of sunspots, {\em Mon.\ Not.\ R. Astr.\ Soc.} {\bf 94}, 39--48 (1934). 

\vspace{1mm} \noi Dudley, M. L., James, R. W. \& Phillips, C. G., Bounds on the infimum decay rate for axisymmetric incompressible dynamos, {\em Geophys.\ Astrophys.\ Fluid Dynam.} {\bf 35}, 373--378 (1986).

\vspace{1mm}\noi Fearn, D.R., Roberts, P.H. \& Soward, A.M., Convection, stability and the dynamo. In: {\em Energy stability and convection}, Pitman Research Notes in Mathematics Series {\bf 168}, Ed. G.P. Galdi \& B. Straughan (Longman Scientific \& Technical, New York 1988), pp.\ 60--324. 

\vspace{1mm} \noi Folland, G.B., {\em Introduction to partial differential equations}, 2nd ed.\ (Princeton Univ.\ Press, Princeton 1995).

\vspace{1mm}\noi Hide, R., The magnetic flux linkage of a moving medium: a theorem and geophysical consequences, {\em J.\ Geophys.\ Res.} {\bf 86}, 11,681--11,687 (1981).

\vspace{1mm} \noi Hide, R. \& Palmer, T.N., Generalization of Cowling's theorem, {\em Geophys.\ Astrophys.\ Fluid Dynam.} {\bf 19}, 301--309 (1982).

\vspace{1mm}\noi Ivers, D. J. \& James, R. W., Axisymmetric antidynamo theorems in compressible nonuniform conducting fluids,
{\em Philos. Trans.\ Roy.\ Soc.\ London Ser.\ A} {\bf 312}, 179--218 (1984).


\vspace{1mm} \noi Kaiser, R., The non-radial velocity theorem revisited, {\em Geophys.\ Astrophys.\ Fluid Dynam.} {\bf 101}, 185--197 (2007).

\vspace{1mm} \noi Kaiser, R., Well-posedness of the kinematic dynamo problem, {\em Math.\ Meth.\ Appl.\ Sci.} {\bf 35}, 1241--1255 (2012).

\vspace{1mm} \noi Kaiser, R., A higher-order-decay result for the dynamo equation with an application to the toroidal velocity theorem (submitted, 2013).

\vspace{1mm}\noi Kaiser, R. \& Uecker, H., Well-posedness of some initial-boundary-value problems for dynamo-generated poloidal magnetic fields, {\em Proc. R. Soc. Edinburgh} {\bf 139A}, 1209--1235 (2009), 
Corrigendum, {\em Proc. R. Soc. Edinburgh} {\bf 141A}, 819--824 (2011), \\
Corrected version, arXiv:1212.3180 [astro-ph.SR] (2012).

\vspace{1mm}\noi Lortz, D., Impossibility of steady dynamos with certain symmetries, {\em Phys.\ Fluids} {\bf 11}, 913--916 (1968).

\vspace{1mm}\noi Lortz, D. \& Meyer-Spasche, R., On the decay of symmetric dynamo fields, {\em Math.\ Meth.\  Appl.\ Sci.} {\bf 4}, 91--97 (1982a).

\vspace{1mm}\noi Lieberman, G.M., {\em Second order parabolic differential equations} (World Scientific, Singapore 1996).

\vspace{1mm}\noi Lortz, D. \& Meyer-Spasche, R., On the decay of symmetric toroidal dynamo fields, {\em Z.\ Naturforsch.} {\bf 37a}, 736--740 (1982b).

\vspace{1mm}\noi Lortz, D., Meyer-Spasche, R. \& Stredulinsky, E. W., Asymptotic Behavior of the Solutions of Certain Parabolic Equations, {\em Comm.\ Pure Appl.\ Math.} {\bf 37}, 677--703 (1984).

\vspace{1mm}\noi Lady\u{z}enskaja, O. A., Solonnikov, V.A. \& Ural'ceva, N. N., {\em Linear and Quasilinear Equations of Parabolic Type}, Translations of Mathematical Monographs Vol.\ {\bf 23} (American Mathematical Society, Providence, R. I. 1968).

\vspace{1mm}\noi Moffatt, H. K., {\em Magnetic Field Generation in Electrically Conducting Fluids} (Cambridge University Press, Cambridge, England 1978).

\vspace{1mm}\noi N$\acute{\rm u}\tilde{\rm n}$ez, M., The decay of axisymmetric magnetic fields: a review of Cowling's theorem, {\em SIAM Review} {\bf 38}, 553--564 (1996). 

\vspace{1mm}\noi Proctor, M.R.E., Homogeneous dynamos. In: {\em Mathematical Aspects of Natural Dynamos}, Ed. E. Dormy \& A.M. Soward (Chapman \& Hall/CRC, Boca Raton, USA 2007), pp.\ 18--41.

\vspace{1mm}\noi Protter, M.H. \& Weinberger, H.R., {\em Maximum Principles in Differential Equations} (Springer, New York 1984).


\vspace{1mm}\noi Stredulinsky, E.W., Meyer-Spasche, R. \& Lortz, D., Asymptotic behavior of solutions of certain parabolic problems
with space and time dependent coefficients, {\em Comm. Pure Appl. Math.} {\bf 39}, 233--266 (1986).

\vspace{1mm}\noi Todoeschuck, J. \& Rochester, M.G., The effect of compressible flow on antidynamo theorems, {\em Nature, Lond.} {\bf 284}, 250--251 (1980).

%
%
\end{document}